\documentclass[reqno,oneside,final]{amsart}
\usepackage[margin=3cm]{geometry}
\usepackage{amssymb,amsmath}
\usepackage{mathtools}
\usepackage{amsfonts}
\usepackage{rotating}
\usepackage{upgreek}
\usepackage{graphicx}
\usepackage{tikz}
\tikzset{
  every overlay node/.style={
    draw=white,fill=white,rounded corners,anchor=north west,
  },
}
\usetikzlibrary{patterns}
\usepackage[siunitx]{circuitikz}

\usepackage{graphics,caption}
\usepackage[labelformat=simple,labelfont=normalfont]{subcaption}

\usepackage[utf8]{inputenc} 
\usepackage[T1]{fontenc} 
\usepackage[boxed,oldcommands,norelsize]{algorithm2e}
\usepackage{hyperref}
\usepackage[numbers,sort&compress]{natbib}
\usepackage{multirow}
\usepackage{hhline}
\usepackage{colortbl}
\newcolumntype{L}{>{\centering\arraybackslash}m{0.46\textwidth}}
\usepackage{arydshln}
\usepackage{comment}

\usepackage{changes}
\usepackage{multirow}

\usepackage{acronym}
\acrodef{fe}[FE]{Finite Element}
\acrodef{dd}[DD]{Domain Decomposition}
\acrodef{bddc}[BDDC]{Balancing Domain Decomposition by Constraints}
\acrodef{dof}[DOF]{Degree Of Freedom}
\acrodefplural{dof}[DOFs]{Degrees of Freedom}
\acrodef{am}[AM]{Additive Manufacturing}
\acrodef{amr}[AMR]{Adaptive Mesh Refinement}
\acrodef{mpi}[MPI]{Message Passing Interface}
\acrodef{hf}[HF]{High Fidelity}
\acrodef{pp}[PP]{Part-Plate}
\acrodef{p}[P]{Part-only}
\acrodef{htc}[HTC]{Heat Transfer Coefficient}
\acrodef{vma}[VMA]{Virtual Mold Approximation}
\acrodef{dmls}[DMLS]{Direct Metal Laser Sintering}
\acrodef{slm}[SLM]{Selective Laser Melting}
\acrodef{vda}[VDA]{Virtual Domain Approximation}
\acrodef{vlp}[VLP]{Virtual Loose Powder}
\acrodef{vp}[VP]{Virtual Plate}
\acrodef{mcam}[MCAM]{Monash Centre for Additive Manufacturing}
\acrodef{pde}[PDE]{Partial Differential Equation}
\acrodef{hpc}[HPC]{High Performance Computing}
\acrodef{pbf}[PBF]{Powder-Bed Fusion}

\newcommand{\T}{\mathrm{T}}

\graphicspath{{Figures/}{Figures/art025/}{Figures/introduction/}{Figures/formulation/}{Figures/experiments/}{Figures/results/}{Figures/verification/}}

\makeatletter
\def\input@path{{Figures/}{Figures/art025/}{Figures/introduction/}{Figures/formulation/}{Figures/experiments/}{Figures/results/}{Figures/verification/}}
\makeatother

\begin{document}

\title[Numerical modelling in PBF with the virtual domain 
approximation]{Numerical modelling of heat transfer and experimental 
validation in Powder-Bed Fusion with the Virtual Domain Approximation}

\author{Eric Neiva$^{\dag,\ddag}$ \and Michele Chiumenti$^{\dag,\ddag}$ \and 
Miguel Cervera$^{\dag,\ddag}$ \and Emilio Salsi$^{\dag,\ddag}$ \and Gabriele 
Piscopo$^{\$}$ \and Santiago Badia$^{\dag,\S}$ \and Alberto F. 
Martín$^{\dag,\ddag}$ \and Zhuoer Chen$^{\P,\pounds}$ \and Caroline Lee$^{\P}$ 
\and Christopher Davies$^{\P}$}

\date{\today}

\renewcommand{\thefootnote}{\arabic{footnote}}

\maketitle

\begin{center}
\small{$^{\dag}$ Centre Internacional de M\`etodes Num\`erics en Enginyeria 
(CIMNE), Building C1, Campus Nord UPC, Gran Capitán S/N, 08034 Barcelona, 
Spain. \{eneiva,michele,mcervera,esalsi,sbadia,amartin\}@cimne.upc.edu.\\}
\end{center}

\begin{center}
\small{$^{\ddag}$ Universitat Polit\`ecnica de Catalunya, Jordi Girona 1-3, 
Edifici C1, 08034 Barcelona, Spain. 
\{eneiva,michele,mcervera,esalsi,sbadia,amartin\}@cimne.upc.edu.\\}
\end{center}

\begin{center}
\small{$^{\$}$ Department of Management and Production Engineering (DIGEP), \\ 
Politecnico di Torino, C.so Duca degli Abruzzi, 24, 10129, Torino, Italy.}
\end{center}

\begin{center}
\small{$^{\S}$ School of Mathematics, Monash University, Clayton, Victoria, 
3800, Australia.}
\end{center}

\begin{center}
\small{$^{\P}$ Department of Mechanical and Aerospace Engineering, Monash 
University, VIC 3800, Australia. \\ Monash Centre for Additive Manufacturing, 
Monash University, VIC 3800, Australia.}
\end{center}

\begin{center}
\small{$^{\pounds}$ Department of Industrial and Materials Science, Chalmers 
University of Technology, 41326 Gothenburg, Sweden.}
\end{center}

\begin{abstract}
Among metal additive manufacturing technologies, powder-bed fusion features 
very thin layers and rapid solidification rates, leading to long build jobs and 
a highly localized process. Many efforts are being devoted to accelerate 
simulation times for practical industrial applications. The new approach 
suggested here, the virtual domain approximation, is a physics-based 
rationale for spatial reduction of the domain in the thermal finite-element 
analysis at the part scale. Computational experiments address, among others, 
validation against a large physical experiment of 17.5 $\mathrm{[cm^3]}$ of 
deposited volume in 647 layers. For fast and automatic parameter estimation at 
such level of complexity, a high-performance computing framework is employed. 
It couples FEMPAR-AM, a specialized parallel finite-element software, with 
Dakota, for the parametric exploration. Compared to previous state-of-the-art, 
this formulation provides higher accuracy at the same computational cost. This 
sets the path to a fully virtualized model, considering an upwards-moving 
domain covering the last printed layers.
\end{abstract}


\pagestyle{myheadings}
\thispagestyle{plain}

\section{Introduction}
\label{sec:introduction}

\ac{am} or 3D Printing is emerging as a prominent 
manufacturing technology \citep{wohlers2017wohlers} in many industrial 
sectors, such as the aerospace, defence, dental or biomedical. These sectors 
are on the lookout to find the way to exploit its many potentials, including 
vast geometrical freedom in design, access to new materials with enhanced 
properties or reduced time-to-market. However, this growth cannot be 
long-term sustained without the support from predictive computer simulation 
tools. Only them provide the appropriate means to jump the hurdle of slow and 
costly trial-and-error physical experimentation, in product design and 
qualification, and to improve the understanding of the
process-structure-property-performance link.

This work concerns the numerical simulation \emph{at the part scale} of metal 
AM processes by \ac{pbf} technologies, such as \ac{dmls}, described in 
\citep[Fig. 1]{chiumenti_neiva_2017}, \ac{slm}, or Electron Beam Melting (EBM). 
Compared to other metal technologies, powder-bed methods feature the thinnest 
layer thicknesses, from 60 to 20 microns (or even below){. As a 
result, building industrial components usually requires depositing thousands of 
layers; thus, from the modelling viewpoint, computational efficiency should be 
at the forefront. Besides, they are also characterized by having fast 
solidification rates. This means that} high heat 
fluxes concentrate in the last printed layers; away from that region, the 
thermal distribution is much smoother and less physically relevant.

Many researchers have used the \ac{fe} method to investigate metal AM 
processes, often aided by their knowledge of modelling other well-known 
processes, such as casting or welding 
\citep{cervera1999thermo,lindgren2014computational}. At the part scale, early 
\ac{fe} models proved their applicability to many engineering problems, e.g. 
selection of process parameters \citep{song_process_2012}, design of scanning 
path \citep{parry_understanding_2016} or evaluation of distortions and 
residual stresses 
\citep{dai2002distortion,prabhakar_computational_2015,lu2018finite}. 
However, numerical tests were often limited to short single-part builds and 
small deposition volumes 
\citep{roberts_three-dimensional_2009,kolossov_3d_2004}.

More recently, the attention has turned to the design of strategies to 
accelerate simulations to tackle longer processes, multiple-part builds, 
higher deposition volumes and, eventually, deal with industrial-scale 
scenarios in reasonable simulation times. Some authors have attempted to 
exploit adaptive mesh refinement
\citep{denlinger_thermal_2016,patil_generalized_2015,KOLLMANNSBERGER20181483} 
and/or parallelization \citep{neiva2018scalable}, while most of them consider 
surrogate models that are inevitably accompanied by some sacrifice of 
accuracy or physical representativeness. 

Typical model simplifications \citep{lindgren2018approaches}, alone or 
combined, consist of time-averaging the history of the process, by lumping 
welds or layers \citep{irwin_line_2016,lindwall2018efficiency}, or reduce the 
domain of analysis by, e.g. excluding the region of loose powder-bed 
surrounding the part and/or the base plate. In the case of thermal analyses, 
heat transfer from the part to the excluded region is then accounted for with 
an equivalent heat loss boundary condition at the solid-powder interface 
\citep{chiumenti_neiva_2017,li2019estimation}. However, determination of these 
boundary conditions has been limited to rather simple approximations and 
challenged by lack of experimental data, especially concerning heat conduction 
through the loose powder. In this sense, a path yet to be explored in metal AM 
is to develop physics-based alternatives, similar to {the \emph{virtual
mould} approach} in the area of casting 
solidification \citep{Dantzig1985,Hong1984,nastac1998monte}, to replace the heat flow model at 
regions of less physical interest with a much faster and less memory demanding 
model.

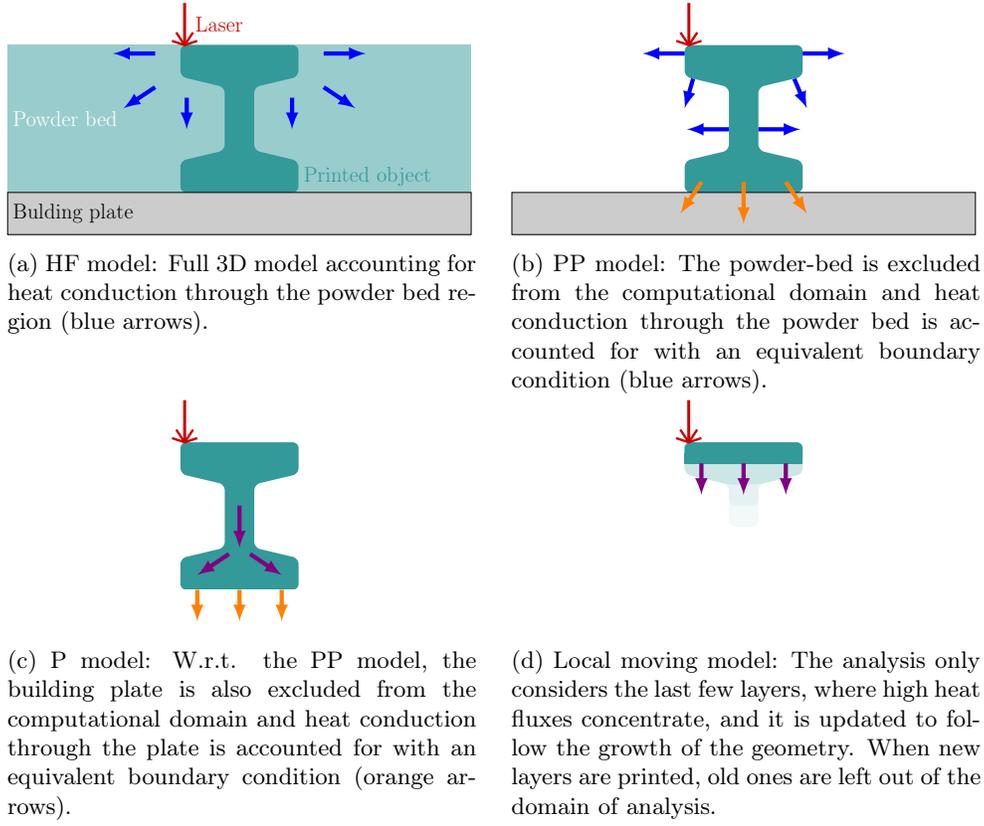
\begin{figure}[!h]
  \centering
  \begin{subfigure}[t]{0.4\textwidth}
    \resizebox{\textwidth}{!}{%
    \begin{tikzpicture}[scale = 1.0]
	
	\fill[fill=teal!40!white] (5, 2.5) -- (16, 2.5) -- (16, 6) -- (5, 6) -- 
	node[text=white,font=\LARGE,anchor=west] {Powder bed} (5, 2.5);
	
	\fill[fill=teal!80!white, rounded corners] (9.1,2.5) -- (9.1,3.25) -- 
	(10.15,3.5) -- (10.15,5.0) -- (9.1,5.25) -- (9.1,6.0) -- (11.9,6.0) -- 
	(11.9,5.25) -- (10.85,5.0) -- (10.85,3.5) -- (11.9,3.25) -- 
	node[text=teal!80!white,font=\LARGE,anchor=west] {Printed object} (11.9,2.5) 
	-- cycle;
	
	\filldraw[fill=gray!40!white, thin] (5, 1.5) -- (16, 1.5) -- (16, 2.5) -- 
	(5,2.5) -- node[font=\LARGE,anchor=west] {Bulding plate} (5,1.5);
	
	\draw[red!80!black, line width = 2.0pt] (9.2, 7) -- 
	node[xshift=0.1cm,font=\LARGE,anchor=west] {Laser} (9.2, 6);
	\draw[red!80!black, ultra thick] (9.2, 6) -- (9.4, 6.3);
	\draw[red!80!black, ultra thick] (8.9, 6.1) -- (9.2, 6) -- (9.5, 6.1);
	\draw[red!80!black, ultra thick] (9.0, 6.3) -- (9.2, 6);
	
	\draw[blue, line width = 1.0mm, -latex] (8.5,5.8) -- (7.5,5.8);
	\draw[blue, line width = 1.0mm, -latex] (12.5,5.8) -- (13.5,5.8);
	
	\draw[blue, line width = 1.0mm, -latex] (8.5,5.0) -- (7.75,4.5);
	\draw[blue, line width = 1.0mm, -latex] (12.5,5.0) -- (13.25,4.5);
	
	\draw[blue, line width = 1.0mm, -latex] (9.25,4.75) -- (9.25,4.0);
	\draw[blue, line width = 1.0mm, -latex] (11.75,4.75) -- (11.75,4.0);
	
\end{tikzpicture}%
    }%
    \caption{\ac{hf} model: Full 3D model accounting for heat conduction 
    through the powder bed region (blue arrows).}
  \end{subfigure} \quad
  \begin{subfigure}[t]{0.4\textwidth}
    \resizebox{\textwidth}{!}{%
    \begin{tikzpicture}[scale = 1.0]
	
	\fill[fill=teal!80!white, rounded corners] (9.1,2.5) -- (9.1,3.25) -- 
	(10.15,3.5) -- (10.15,5.0) -- (9.1,5.25) -- (9.1,6.0) -- (11.9,6.0) -- 
	(11.9,5.25) -- (10.85,5.0) -- (10.85,3.5) -- (11.9,3.25) -- (11.9,2.5) -- 
	cycle;
	
	\filldraw[fill=gray!40!white, thin] (5, 1.5) -- (16, 1.5) -- (16, 2.5) -- (5, 2.5) -- cycle;
	
	\draw[red!80!black, line width = 2.0pt] (9.2, 7) -- (9.2, 6);
    \draw[red!80!black, ultra thick] (9.2, 6) -- (9.4, 6.3);
    \draw[red!80!black, ultra thick] (8.9, 6.1) -- (9.2, 6) -- (9.5, 6.1);
    \draw[red!80!black, ultra thick] (9.0, 6.3) -- (9.2, 6);
    
	\draw[blue, line width = 1.0mm, -latex] (9.1,5.8) -- (8.1,5.8);
	\draw[blue, line width = 1.0mm, -latex] (11.9,5.8) -- (12.9,5.8);
	
	\draw[blue, line width = 1.0mm, -latex] (9.3,5.2) -- (9.1,4.5);
	\draw[blue, line width = 1.0mm, -latex] (11.7,5.2) -- (12.0,4.5);
	
	\draw[blue, line width = 1.0mm, -latex] (10.15,4.0) -- (9.15,4.0);
	\draw[blue, line width = 1.0mm, -latex] (10.85,4.0) -- (11.85,4.0);
	
	\draw[orange, line width = 1.0mm, -latex] (10.5,2.75) -- (10.5,1.75);
	\draw[orange, line width = 1.0mm, -latex] (11.5,2.75) -- (12.0,2.00);
	\draw[orange, line width = 1.0mm, -latex] (9.50,2.75) -- (9.00,2.00);	
			
\end{tikzpicture}%
    }%
    \caption{\ac{pp} model: The powder-bed is excluded from the 
    computational domain and heat conduction through the powder bed is 
    accounted for with an equivalent boundary condition (blue arrows).}
  \end{subfigure} \\
  \begin{subfigure}[t]{0.4\textwidth}
    \resizebox{\textwidth}{!}{%
    \begin{tikzpicture}[scale = 1.0]
	
	\fill[fill=teal!80!white, rounded corners] (9.1,2.5) -- (9.1,3.25) -- 
	(10.15,3.5) -- (10.15,5.0) -- (9.1,5.25) -- (9.1,6.0) -- (11.9,6.0) -- 
	(11.9,5.25) -- (10.85,5.0) -- (10.85,3.5) -- (11.9,3.25) -- (11.9,2.5) -- 
	cycle;
	
	\fill[white] (5, 1.5) -- (16, 1.5) -- (16, 2.5) -- (5, 2.5) -- cycle;
	
	\draw[red!80!black, line width = 2.0pt] (9.2, 7) -- (9.2, 6);
    \draw[red!80!black, ultra thick] (9.2, 6) -- (9.4, 6.3);
    \draw[red!80!black, ultra thick] (8.9, 6.1) -- (9.2, 6) -- (9.5, 6.1);
    \draw[red!80!black, ultra thick] (9.0, 6.3) -- (9.2, 6);
    
	
	
	
	\draw[orange, line width = 1.0mm, -latex] (10.5,2.5) -- (10.5,1.75);
	\draw[orange, line width = 1.0mm, -latex] (11.5,2.5) -- (11.5,1.75);
	\draw[orange, line width = 1.0mm, -latex] (9.50,2.5) -- (9.50,1.75);	
	
	\draw[violet, line width = 1.0mm, -latex] (10.5,4.50) -- (10.50,3.50);
	\draw[violet, line width = 1.0mm, -latex] (10.75,3.35) -- (11.5,2.85);
	\draw[violet, line width = 1.0mm, -latex] (10.25,3.35) -- (9.50,2.85);
			
\end{tikzpicture}%
    }%
    \caption{\ac{p} model: W.r.t. the \ac{pp} model, the building plate is 
    also excluded from the computational domain and heat conduction through 
    the plate is accounted for with an equivalent boundary condition (orange 
    arrows).}
  \end{subfigure} \quad
  \begin{subfigure}[t]{0.4\textwidth}
    \resizebox{\textwidth}{!}{%
    \begin{tikzpicture}[scale = 1.0]
	

	\fill[fill=teal!5!white, rounded corners] (10.15,4.0) -- (10.15,5.0) -- 
	(9.1,5.25) -- (9.1,6.0) -- (11.9,6.0) -- (11.9,5.25) -- (10.85,5.0) -- 
	(10.85,4.0) -- cycle;
	\fill[fill=teal!10!white, rounded corners] (10.15,4.5) -- (10.15,5.0) -- 
	(9.1,5.25) -- (9.1,6.0) -- (11.9,6.0) -- (11.9,5.25) -- (10.85,5.0) -- 
	(10.85,4.5) -- cycle;
	\fill[teal!20!white, rounded corners] (10.15,5.0) -- 
	(9.1,5.25) -- (9.1,6.0) -- (11.9,6.0) -- (11.9,5.25) -- (10.85,5.0) -- cycle;
	\fill[teal!80!white, rounded corners] (9.1,5.5) -- (9.1,6.0) -- (11.9,6.0) -- 
	(11.9,5.5) -- cycle;
	\fill[teal!80!white] (9.1,5.5) -- (9.1,5.75) -- (11.9,5.75) -- 
	(11.9,5.5) -- cycle;
	
	\fill[white] (5, 1.5) -- (16, 1.5) -- (16, 2.5) -- (5, 2.5) -- cycle;
	
	\draw[red!80!black, line width = 2.0pt] (9.2, 7) -- (9.2, 6);
    \draw[red!80!black, ultra thick] (9.2, 6) -- (9.4, 6.3);
    \draw[red!80!black, ultra thick] (8.9, 6.1) -- (9.2, 6) -- (9.5, 6.1);
    \draw[red!80!black, ultra thick] (9.0, 6.3) -- (9.2, 6);
    
	
	
	
	\draw[violet, line width = 1.0mm, -latex] (10.5,5.5) -- (10.5,4.75);
	\draw[violet, line width = 1.0mm, -latex] (11.5,5.5) -- (11.5,4.75);
	\draw[violet, line width = 1.0mm, -latex] (9.50,5.5) -- (9.50,4.75);
			
\end{tikzpicture}%
    }%
    \caption{Local moving model: The analysis only considers the 
    last few layers, where high heat fluxes concentrate, and it is updated to 
    follow the growth of the geometry. When new layers are printed, old ones 
    are left out of the domain of analysis.}
    \label{fig:methodology_goal}
  \end{subfigure}
  \caption{Application of the \ac{vda} technique. The reference \ac{hf} model 
  includes all the regions involved in the process, but relevant heat transfer 
  in powder-bed methods occurs only at the last printed layers. According to 
  this, the powder-bed, the building plate and previous layers 
  of the solid part can be removed from the computational domain and 
  heat conduction through them can be accounted for with equivalent boundary 
  conditions (blue for powder-bed, orange for plate and violet for part).}
  \label{fig:overview_methodology}
\end{figure}

The purpose of this work is to establish a new \emph{physics-based} rationale 
for domain reduction in the thermal \ac{fe} analysis of metal AM processes by 
powder-bed methods. The new technique, referred to as the \ac{vda} in 
Sect.~\ref{sec:virtual}, approximates the 3D transient heterogeneous heat 
flow problem (see Sect.~\ref{sec:formulation}) at low-relevance subregions 
(e.g. loose powder, building plate) by a 1D heat conduction problem (see Fig. 
\ref{fig:overview_methodology}). This much simpler 1D problem can be 
analytically solved and reformulated as an equivalent boundary condition for 
the 3D reduced-domain problem.

Applying this formulation to a simple proof-of-concept example in Sect. 
\ref{sec:verification}, reduced \ac{pp} and \ac{p} models are derived and 
compared with respect to a full powder-bed-part-plate \ac{hf} model.  
{In the results of this example,} 
the spatially reduced models {are able to accurately approximate the 
reference response with a computational runtime more than ten times lower than 
the full model}. A second numerical test in Sect.~\ref{sec:results} addresses 
experimental validation against physical tests carried out at the \ac{mcam} in 
Melbourne, Australia, using an EOSINT M280 machine and Ti-6Al-4V powder. In 
this case, a \ac{hf} model is first calibrated and validated against the 
experimental data. This step is next repeated for two ancillary \ac{pp} reduced 
models. In the first one, the \ac{htc} between the part and the powder is 
assumed constant, as in earlier works 
\citep{chiumenti_neiva_2017,chiumenti_numerical_2017}. In the second one, the 
\ac{pp} adopts the \ac{vda} formulation.

Given the scale of the experiment, 17.5 $\mathrm{[cm^3]}$ of deposited volume 
in 647 layers and 3.5 $\mathrm{[h]}$ of process, the \ac{hf} model size 
amounts to 9.8 million unknowns, whereas the \ac{pp} models to 0.7 million. 
Sensitivity analysis and parameter estimation at such level of complexity are 
only practical by means of an advanced computational framework. For this 
reason, the numerical tests are supported by a high-end parallel computing 
framework, unprecedented in the simulation of metal AM processes. The 
framework combines three tools: (1) FEMPAR-AM \citep{neiva2018scalable}, a 
module of \href{http://www.fempar.org/}{FEMPAR}~\cite{badia2017fempar}, a 
general-purpose object-oriented message-passing/multi-threaded 
scientific software for the fast solution of multiphysics problems governed 
by \acp{pde}; (2) \href{https://dakota.sandia.gov/}{Dakota}~\cite{dakota}, for 
the automatic parametric exploration of the models; and (3) 
\href{https://caminstech.upc.edu/es/calculintensiu}{TITANI}~\ref{tab:titani}, 
a \ac{hpc} cluster at the Universitat Politècnica de 
Catalunya (Barcelona, Spain) to support the calculations. Using this 
innovative methodology for the MCAM experiment, the physics-based 
\ac{vda}-\ac{pp} model is shown to reproduce the response of the \ac{hf} model 
with increased accuracy, with respect to the constant \ac{htc}-\ac{pp} variant. 
However, the simulation times of both reduced models is practically the same. 
In other words, the extra cost devoted to evaluate the equivalent boundary 
condition with the \ac{vda} formulation is negligible, in front of the 
computational cost devoted to the solution of the problem at the reduced 
domain. 

As a result, the \ac{vda} formulation offers good compromise between accuracy 
and efficiency. Indeed, on the one hand, the computational benefit is clear, 
as the mesh covers a smaller region and the number of degrees of freedom with 
respect to the complete model is significantly reduced. On the other hand, 
the impact on accuracy is efficiently controlled, because the neglected 
physics are taken into account in the evaluation of the equivalent heat loss 
boundary condition, without affecting the computational cost of the 
simulation. Even though the \ac{vda} formulation does not totally get rid of 
cumbersome parameter estimations, it can be easily calibrated with respect to 
a \ac{hf} model. This turns it into an appealing alternative for sequential 
coupling with a part-scale mechanical simulation or optimization problems 
(e.g. design of minimum-distortion scanning path). More interestingly, by 
exploiting the locality of the \ac{pbf} process, the \ac{vda} could eventually 
be useful to reduce the domain down to a few layers, as shown in Fig. 
\ref{fig:methodology_goal}, while keeping good relation with the thermal 
response of the \ac{hf} model.

\section{Heat transfer analysis in AM}
\label{sec:formulation}

This section describes the transient model for the thermal \ac{fe} analysis
of the printing process \emph{at the part scale}. The contents centre upon (1) 
an overview of the model variants studied in this work, which depend on the
domain of analysis and heat loss model (Sect. \ref{sec:variants}); (2) the 
governing equation and the determination of the material properties at the
powder state (Sect. \ref{sec:governing}); and (3) the treatment of boundary
conditions, according to the heat loss model (Sect. \ref{sec:boundary}). The
reader is referred to~\citep{chiumenti_numerical_2017,chiumenti_neiva_2017} 
for further details on the weak formulation and the \ac{fe} modelling of the 
geometry growth during the metal deposition process. 

\subsection{Model variants}
\label{sec:variants}

Let $\Omega^{\rm pbf}$ be an open bounded domain in $\mathbb{R}^3$,
representing the system formed by the printed object $\Omega^{\rm part}$, the
building plate $\Omega^{\rm base}$ and the surrounding powder bed
$\Omega^{\rm bed}$. $\Omega^{\rm pbf}$ grows in time during the build
process. After the printing, it remains fixed, while cooling down to room
temperature. 

\begin{figure}[!h]
  \centering
  \begin{subfigure}[t]{0.3\textwidth}
    \resizebox{\textwidth}{!}{%
    \begin{tikzpicture}[scale = 1.0]
	
	\fill[fill=teal!40!white] (5, 2.5) -- (10.5, 2.5) -- (10.5, 6) -- (5, 6) 
	-- cycle;
	
	\fill[fill=teal!80!white, rounded corners] (9.1,2.5) -- (9.1,3.25) -- 
	(10.15,3.5) -- (10.15,5.0) -- (9.1,5.25) -- (9.1,6.0) -- (10.5,6.0) -- 
	(10.5,2.5) -- cycle;
	
	\fill[fill=teal!80!white] (10.5,6.0) rectangle (10.4,2.5);
	
	\fill[gray!40!white] (5, 1.5) -- (10.5, 1.5) -- 
	(10.5, 2.5) -- (5, 2.5) -- cycle;
	\draw[thin] (10.5, 2.5) -- (5, 2.5) -- (5, 1.5) -- (10.5, 1.5);
	
	\node[font=\LARGE, anchor=west] at (9.2,3.0) {$\Omega^{\rm part}$};
	\node[font=\LARGE, anchor=west] at (5.5,2.0) {$\Omega^{\rm base}$};
	\node[font=\LARGE, anchor=west] at (5.5,4.0) {$\Omega^{\rm bed}$};
	
    \draw[olive, ultra thick] (5.0, 1.5) -- node[font=\LARGE, anchor = north] 
    {$\partial \Omega_{\rm down}$} (10.5, 1.5);
    \draw[blue, dashed, ultra thick] (5.0, 2.5) -- node[font=\LARGE, anchor = 
    east] {$\partial \Omega_{\rm lat}^{\rm bed}$} (5.0, 6.0);
    \draw[blue, ultra thick] (5.0, 1.5) -- node[font=\LARGE, anchor = east] 
    {$\partial \Omega_{\rm lat}^{\rm base}$} (5.0, 2.5);
    \draw[orange, dotted, ultra thick] (5.0, 6.0) -- node[font=\LARGE, anchor 
    = south] {$\partial \Omega_{\rm air}^{\rm bed}$} (9.2, 6.0);
    \draw[orange, ultra thick] (9.2, 6.0) -- node[font=\LARGE, anchor 
    = south, xshift = -0.25cm] {$\partial \Omega_{\rm air}^{\rm part}$} 
    (10.5, 6.0);
    
    \draw[gray, dashdotted, line width = 1mm] (10.5,1.0) -- node[pos = 0.95, 
    anchor = east] {sym.} (10.5,7.5);
			
\end{tikzpicture}%
    }%
    \caption{\ac{hf} model.}
  \end{subfigure} \
  \begin{subfigure}[t]{0.3\textwidth}
    \resizebox{\textwidth}{!}{%
    \begin{tikzpicture}[scale = 1.0]
	
	\fill[fill=teal!80!white, rounded corners] (9.1,2.5) -- (9.1,3.25) -- 
	(10.15,3.5) -- (10.15,5.0) -- (9.1,5.25) -- (9.1,6.0) -- (10.5,6.0) -- 
	(10.5,2.5) -- cycle;
	
	\fill[gray!40!white] (5, 1.5) -- (10.5, 1.5) -- 
	(10.5, 2.5) -- (5, 2.5) -- cycle;
	\draw[thin] (10.5, 2.5) -- (5, 2.5) -- (5, 1.5) -- (10.5, 1.5);
	
	\fill[fill=teal!80!white] (10.5,6.0) rectangle (10.4,2.5);
	
	\node[font=\LARGE, anchor=west] at (9.2,3.0) {$\Omega^{\rm part}$};
	\node[font=\LARGE, anchor=west] at (5.5,2.0) {$\Omega^{\rm base}$};
	
    \draw[olive, ultra thick] (5.0, 1.5) -- node[font=\LARGE, anchor = north] 
    {$\partial \Omega_{\rm down}$} (10.5, 1.5);
    \draw[blue, ultra thick] (5.0, 1.5) -- node[font=\LARGE, anchor = east] 
    {$\partial \Omega_{\rm lat}^{\rm base}$} (5.0, 2.5);
    \draw[orange, ultra thick] (9.1, 6.0) -- node[font=\LARGE, anchor 
    = south, xshift = -0.25cm] {$\partial \Omega_{\rm air}^{\rm part}$} 
    (10.5, 6.0);
    \draw[red, rounded corners, ultra thick] (9.1,2.5) -- (9.1,3.25) -- 
    (10.15,3.5) -- node[font=\LARGE, anchor = east] {$\partial \Omega_{\rm 
    bed}^{\rm part}$} 
    (10.15,5.0) -- (9.1,5.25) -- (9.1,6.0);
    \draw[red, dashdotted, ultra thick] (5.0,2.5) -- node[font=\LARGE, anchor = 
    south] {$\partial \Omega_{\rm bed}^{\rm base}$} (9.1,2.5);
    
    \draw[gray, dashdotted, line width = 1mm] (10.5,1.0) -- node[pos = 0.95, 
    anchor = east] {sym.} (10.5,7.5);
			
\end{tikzpicture}%
    }%
    \caption{\ac{pp} model.}
  \end{subfigure} \
  \begin{subfigure}[t]{0.3\textwidth}
    \resizebox{\textwidth}{!}{%
    \begin{tikzpicture}[scale = 1.0]
	
	\fill[fill=teal!80!white, rounded corners] (9.1,2.5) -- (9.1,3.25) -- 
	(10.15,3.5) -- (10.15,5.0) -- (9.1,5.25) -- (9.1,6.0) -- (10.5,6.0) -- 
	(10.5,2.5) -- cycle;
	
	\fill[white] (5, 1.5) -- (10.5, 1.5) -- 
	(10.5, 2.5) -- (5, 2.5) -- cycle;
	\draw[white] (10.5, 2.5) -- (5, 2.5) -- (5, 1.5) -- (10.5, 1.5);
	
	\fill[fill=teal!80!white] (10.5,6.0) rectangle (10.4,2.5);
	
	\node[font=\LARGE, anchor=west] at (9.2,3.0) {$\Omega^{\rm part}$};
	
    \draw[white, ultra thick] (5.0, 1.5) -- node[font=\LARGE, anchor = east] 
    {$\partial \Omega_{\rm lat}^{\rm base}$} (5.0, 2.5);
    \draw[orange, ultra thick] (9.1, 6.0) -- node[font=\LARGE, anchor 
    = south, xshift = -0.25cm] {$\partial \Omega_{\rm air}^{\rm part}$} 
    (10.5, 6.0);
    \draw[red, rounded corners, ultra thick] (9.1,2.5) -- (9.1,3.25) -- 
    (10.15,3.5) -- node[font=\LARGE, anchor = east] {$\partial \Omega_{\rm 
    bed}^{\rm part}$} 
    (10.15,5.0) -- (9.1,5.25) -- (9.1,6.0);
    \draw[violet, ultra thick] (9.1, 2.5) -- node[font=\LARGE, anchor = north, 
    xshift = -0.25cm] {$\partial \Omega_{\rm base}^{\rm part}$} (10.5, 2.5);
    
    \draw[gray, dashdotted, line width = 1mm] (10.5,1.0) -- node[pos = 0.95, 
    anchor = east] {sym.} (10.5,7.5);
		
\end{tikzpicture}%
    }%
    \caption{\ac{p} model.}
  \end{subfigure}
  \caption{Close-up of Fig.~\ref{fig:overview_methodology} to illustrate the 
  thermal model variants studied in this work, along with the boundary 
  conditions.}
  \label{fig:overview_models}
\end{figure}
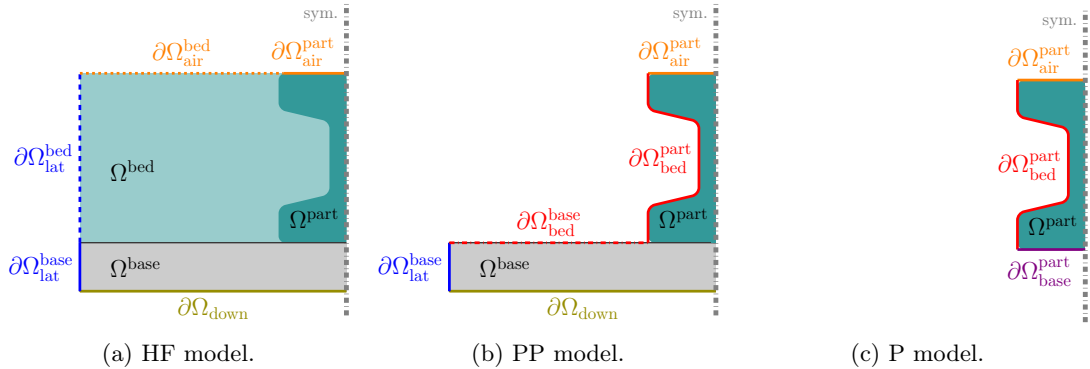

Several variants of the thermal model, represented in 
Fig.~\ref{fig:overview_models}, arise from considering different 
computational domains $\Omega$:
\begin{enumerate}
	\item If $\Omega \equiv \Omega^{\rm pbf} = \Omega^{\rm part} \cup
	\Omega^{\rm base} \cup \Omega^{\rm bed}$, the model is referred to as
	\textbf{\ac{hf}}. The contour of $\Omega$ is formed by a region in 
	contact with the air in the chamber $\partial \Omega_{\rm air}^{\rm bed} 
	\cup \partial \Omega_{\rm air}^{\rm part}$, the lateral wall $\partial 
	\Omega_{\rm lat}^{\rm bed} \cup \Omega_{\rm lat}^{\rm base}$ and the 
	bottom wall of the plate $\partial \Omega_{\rm down}$.
	\item If the powder bed is excluded from the computational domain, i.e.,
	$\Omega = \Omega^{\rm part} \cup \Omega^{\rm base}$, then the 
	so-called \textbf{\ac{pp}} model is obtained. In this case, the solid-powder
	interface is also in the boundary, i.e., $\partial \Omega = 
	\partial \Omega_{\rm air}^{\rm part} \cup \partial \Omega_{\rm bed}^{\rm
	part} \cup \partial \Omega_{\rm bed}^{\rm base} \cup \partial \Omega_{\rm 
	lat}^{\rm base} \cup \partial \Omega_{\rm down}$.
	\item Finally, taking only $\Omega \equiv \Omega^{\rm part}$ yields the
	\textbf{\ac{p}} model, with $\partial \Omega = \partial \Omega_{\rm air}^{\rm
	part} \cup \partial \Omega_{\rm bed}^{\rm part} \cup \partial \Omega_{\rm
	base}^{\rm part}$.
\end{enumerate}

From the computational viewpoint, \ac{hf} is the most demanding model, 
because the part, the plate and the powder bed must be meshed, whereas 
\ac{p} is the simplest, because it only needs to mesh the part. On the other
hand, \ac{hf} is the most physically representative model, among those
studied here. Therefore, it is established as the \emph{reference} numerical 
model; i.e. all reduced variants are compared against this one.

As a result of excluding the powder bed (and the building plate), heat
conduction through the powder (and the plate) must be accounted for with
an equivalent heat loss boundary condition at $\partial \Omega_{\rm bed}^{\rm 
part} \cup \partial \Omega_{\rm bed}^{\rm base}$ (or $\partial \Omega_{\rm 
bed}^{\rm part}$ and $\partial \Omega_{\rm base}^{\rm part}$). For this 
purpose, two strategies are studied: (1) constant-valued \ac{htc} and (2) 
\ac{vda} (cf. Sect. \ref{sec:virtual}). They lead to further submodels, such as 
\ac{htc}-\ac{pp}, \ac{vda}-\ac{pp} and \ac{vda}-\ac{p}. Only these three are 
covered in this work. In particular, their computational cost and accuracy with 
respect to the reference \ac{hf} is assessed in Sect. \ref{sec:numerical}.

\subsection{Governing equation}
\label{sec:governing}

Heat transfer in $\Omega$ at the part scale is governed by the balance of
energy equation, expressed as
\begin{equation}
    C(T) \partial_{t}{T} - \boldsymbol{\nabla} \cdot ( k(T) \nabla T ) = r, 
    \quad \text{in} \enspace \Omega(t), \quad t > 0,    
\label{eq:balanceofenergy}
\end{equation}
where $C(T)$ is the heat capacity coefficient, given by the product of the 
density of the material $\rho(T)$ and the specific heat $c(T)$, and $k(T) \geq 
0$ is the thermal conductivity. Furthermore, $r$ is the rate of energy 
supplied to the system per unit volume by a very intense and concentrated 
laser or electron beam that moves in time according to a user-defined 
deposition sequence, referred to as the \emph{scanning path}. \added{After 
integrating Eq. \eqref{eq:balanceofenergy} in time,} $r$ is computed as 
\begin{equation}
	\begin{cases}
		r\added{(\boldsymbol{x})} = \frac{\eta W}{\mathrm{V}_{\rm 
		pool}^{\mathrm{\Delta \mathrm{t}}}} & \added{\text{if } \boldsymbol{x} \in 
		\mathrm{V}_{\rm pool}^{\mathrm{\Delta \mathrm{t}}}} \\
		\added{0} & \added{\text{elsewhere}} \\
	\end{cases}
\end{equation}
with $W$ the laser power $\mathrm{[W]}$, $\eta$ the heat absorption 
coefficient, a measure of the laser efficiency, and $\mathrm{V}_{\rm 
pool}^{\mathrm{\Delta \mathrm{t}}}$ is the region swept by the laser during the 
time increment $\Delta \mathrm{t}$. Note that phase transformations occur much 
faster than the diffusion process and the amount of latent heat is much smaller 
than the energy input~\citep{chiumenti_neiva_2017}. That is why, given the 
scale of analysis, phase-change effects are neglected.

In case of the \ac{hf} model, the powder-bed is included into the 
computational domain. As a result, there are two distinct material phases 
playing a role in Eq. \eqref{eq:balanceofenergy}, i.e. solid and powder, which 
are separated at the $\partial \Omega_{\rm bed}^{\rm part} \cup \partial 
\Omega_{\rm bed}^{\rm base}$ interface between the solid part-plate ensemble 
and the granular powder-bed. To model the material in powder state, the 
thermophysical properties are determined in terms of the bulk material data 
at solid state and the porosity of the powder-bed $\phi$. In particular, 
density and specific heat are obtained as
\begin{align}
    \rho_{\rm pwd} &= \rho_{\rm solid} ( 1 - \phi ), \quad \text{and}\\
    c_{\rm pwd} &= c_{\rm solid},
\end{align}
\noindent{while determination of thermal conductivity $k_{\rm pwd}$ is 
more involved and often relies on empirical expressions. Among the models 
present in the literature, Sih and Barlow~\citep{sih2004prediction} establish,
for a powder-bed composed of spherical particles
\citep{denlinger_thermal_2016}, the relation}
\begin{equation}
	\begin{aligned}
    \frac{k_{\rm pwd}}{k_{\rm gas}} &= \left( 1 - \sqrt{1-\phi} \right) 
    \left( 1 + \phi \frac{k_{\rm rad}}{k_{\rm gas}} \right) \\ &+ \sqrt{1-\phi} 
    \frac{2}{1-\frac{k_{\rm gas}}{k_{\rm solid}}} \left( 
    \frac{2}{1-\frac{k_{\rm gas}}{k_{\rm solid}}} \ln \frac{k_{\rm 
    solid}}{k_{\rm gas}} - 1 \right) \\ &+ \sqrt{1-\phi} \frac{k_{\rm 
    rad}}{k_{\rm gas}} 
	\end{aligned}
\label{eq:powderconductivity}
\end{equation}
\noindent{where $k_{\rm gas}$ is the thermal conductivity of the surrounding 
air or gas and $k_{\rm rad}$ is the contribution of radiation amongst the 
individual powder particles, given by Damköhler's equation:}
\begin{equation}
    k_{\rm rad} = \frac{4}{3} \sigma T^3 D_{\rm pwd}, 
\label{eq:damkohler}
\end{equation}
\noindent{with $D_{\rm pwd}$ the average diameter of the particles and 
$\sigma$ the Stefan-Boltzmann constant, i.e. 
$5.67 \cdot 10^{-8} \ [\mathrm{W} \mathrm{m}^{-2} \mathrm{K}^{-4}]$.}

\subsection{Boundary conditions}
\label{sec:boundary}

Eq.~\eqref{eq:balanceofenergy} is subject to the initial condition
\begin{equation}
	T(\boldsymbol{x},0) = T_0,
\label{eq:initial condition}
\end{equation}
\noindent{with $T_0$ the pre-heating temperature of the build chamber, and 
the boundary conditions applied on the regions shown in 
Fig.~\ref{fig:overview_models}, which depend on the model variant:}
\begin{enumerate}
	\item \emph{\ac{hf} model:} Heat convection and radiation through the 
	free surface $\partial \Omega_{\rm air}^{\rm bed} \cup \partial 
	\Omega_{\rm air}^{\rm part}$, heat conduction through the lateral wall on 
	$\partial \Omega_{\rm lat}^{\rm bed} \cup \partial \Omega_{\rm lat}^{\rm 
	base}$ and heat conduction through the bottom wall of the plate on
	$\partial \Omega_{\rm down}$.
	\item \emph{\ac{pp} model:} Heat convection and radiation only through 
	$\partial \Omega_{\rm air}^{\rm part}$, heat conduction through the 
	powder bed along $\partial \Omega_{\rm bed}^{\rm part} \cup \partial
	\Omega_{\rm bed}^{\rm base}$, heat conduction through the lateral wall
	only on $\partial \Omega_{\rm lat}^{\rm base}$ and heat conduction
	through the bottom wall of the plate on $\partial \Omega_{\rm down}$.
	\item \emph{\ac{p} model:} Heat convection and radiation through 
	$\partial \Omega_{\rm air}^{\rm part}$, heat conduction through the 
	powder bed along $\partial \Omega_{\rm bed}^{\rm part}$ and heat 
	conduction through the building plate on $\partial \Omega_{\rm
	base}^{\rm part}$.
\end{enumerate}

After linearising the Stefan-Boltzmann law for heat radiation
\citep{chiumenti_numerical_2017}, all heat loss boundary conditions
mentioned above can be expressed in terms of the Newton law of cooling:
\begin{equation}
    q_{\rm loss}(T,t) = h_{\rm loss}(T) (T - T_{\rm loss}(t)),
\label{eq:bondary_conditions}
\end{equation}
\noindent{in $\partial \Omega^{\rm loss}(t)$, $t > 0$, where $loss$ refers to 
the kind of heat loss mechanism and the boundary region where it applies. 
Alternatives to Eq.~\eqref{eq:bondary_conditions}} can also be considered, e.g. 
the temperature on $\partial \Omega_{\rm down}$ can be prescribed to $T_0$
\citep{chiumenti_neiva_2017}, taking into account that thermal inertia at
the building plate is much larger than at the printed part.

For the \ac{htc} model, $h_{\rm loss}$ and $T_{\rm loss}$ are always taken as 
constant, due to the difficulty in describing experimentally the temperature 
dependency of these quantities. On the other hand, for heat loss through a 
region modelled with a \ac{vda}, $h_{\rm loss}$ and $T_{\rm loss}$ are both 
temperature dependent and computed as described in Sect.~\ref{sec:virtual}.

\section{Virtual domain approximation}
\label{sec:virtual}

As mentioned in Sect.~\ref{sec:variants}, if the powder-bed (and the building 
plate) are not included in the domain of analysis, then heat transfer 
through these regions must be accounted for with equivalent heat conduction 
boundary conditions. According to Eq.~\eqref{eq:bondary_conditions}, this 
leads to the determination of $h_{\rm loss}$ and $T_{\rm loss}$ values for
heat loss through the solid-powder interface and the part-plate interface.

However, lack of experimental data and physical modelling in the literature 
are a hurdle in the way to estimate these quantities. The authors are not 
aware of the existence of, for instance, well-established
temperature-dependent empirical correlations for the \ac{htc} solid-powder or
approximate time evolution laws for the temperature at the building plate.

To overcome these challenges, this work introduces a novel method, i.e. the 
Virtual Domain Approximation (\ac{vda}). The \ac{vda} enhances a 
state-of-the-art
thermal contact model for metal casting analysis \cite{chiumenti2008numerical},
such that it includes the effect of the thermal inertia of the powder bed (or 
the building plate). The result of the \ac{vda} procedure is a 
temperature-dependent physics-based way to evaluate $h_{\rm loss}$ and $T_{\rm
loss}$ in Eq.~\eqref{eq:bondary_conditions}, a clear improvement with respect 
to the \ac{htc} model.

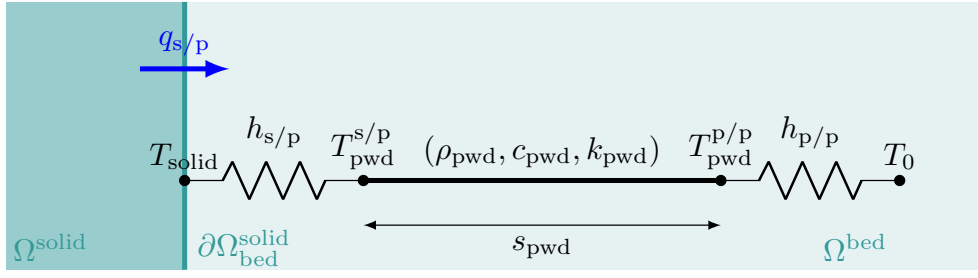
\begin{figure}[!h]
  \centering
  \resizebox{0.85\textwidth}{!}{%


\begin{tikzpicture}
	\fill[teal!40!white] (-2,-1) rectangle (0,2);
	\fill[teal!10!white] (0,-1) rectangle (9,2);
	\draw[teal!80!white, ultra thick] (0,2) -- node[pos=0.917, anchor = west]
	{$\partial \Omega_{\rm bed}^{\rm solid}$} (0,-1);
	\node[teal!80!white] at (-1.5,-0.75) {$\Omega^{\rm solid}$};
	\node[teal!80!white] at (7.5,-0.75) {$\Omega^{\rm bed}$};
    \newcommand{\powder}{$(\rho_{\rm pwd},c_{\rm pwd},k_{\rm pwd})$}
    \draw[color=black] (0,0) to [R, l=$h_{\rm s/p}$,*-*] (2,0);
    \draw[color=black] (6,0) to [R, l=$h_{\rm p/p}$,*-*] (8,0);
    \draw[color=black,ultra thick] (2,0) -- node[above] {\powder} (6,0);
    \draw[color=black,latex-latex] (2,-0.5) -- (6,-0.5) node[midway,below] 
    {$s_{\rm pwd}$};
    \node[above] at (0,0) {$T_{\rm solid}$};
    \node[above] at (2,0) {$T_{\rm pwd}^{\rm s/p}$};
    \node[above] at (6,0) {$T_{\rm pwd}^{\rm p/p}$};
    \node[above] at (8,0) {$T_0$};
    \draw[blue,ultra thick,-latex] (-0.5,1.25) -- node[anchor = south] {$q_{\rm 
    s/p}$} (0.5,1.25);
\end{tikzpicture}%
  }%
  \caption{Illustration of the \ac{vda} method. The main assumption of the 
  method consists in replacing the 3D heat conduction problem across 
  $\Omega^{\rm bed}$ by an equivalent 1D heat conduction problem with the 
  thermal circuit shown in the figure. Then, Eq.~\ref{eq:balanceofenergy} is 
  only solved in $\Omega^{\rm solid}$. From the solution of the equivalent 1D 
  problem a boundary condition of the form given in 
  Eq.~\ref{eq:bondary_conditions} (with $\rm loss = \rm s/p$) is evaluated at 
  each integration point located on $\partial \Omega_{\rm bed}^{\rm solid}$.}
  \label{fig:vda_thermal_circuit}
\end{figure}

Assuming the \ac{pp} variant, the \ac{vda} method is outlined in 
Fig.~\ref{fig:vda_thermal_circuit} and explained in the following paragraphs. 
It is based on the assumption that the entire 3D heat transfer problem across 
the powder bed $\Omega^{\rm bed}$ can be modelled with an equivalent 1D heat 
conduction problem across \replaced{a wall on the contour of}{a 
contouring wall on} $\partial \Omega_{\rm bed}^{\rm part} \cup\partial 
\Omega_{\rm bed}^{\rm base}$. \replaced{As a result, heat loss through the 
powder bed is assumed to be unidimensional and orthogonal to the solid-powder 
interface, i.e. the \ac{vda} neglects diffusion in directions other than the 
normal to the interface.}{This problem is, in fact, equivalent to the classic 
1D transient heat transfer through a plane wall 
\citep{bergman2011fundamentals}. Note that heat loss is not only assumed to be 
unidimensional, but also \emph{orthogonal} to the element face on the 
\emph{discretized} solid-powder interface, i.e. the \ac{vda} neglects diffusion 
in directions other than the normal to the boundary face.}

\added{According to this, the setting considers the 1D time-dependent heat 
transfer problem through the powder bed given by}
\begin{equation}
    \added{\rho_{\rm pwd}(T) c_{\rm pwd}(T) \partial_{t}{T} - \partial_{x} ( 
    k_{\rm pwd}(T) \partial_{x} T ) = 0, \quad \text{in} \enspace (0,s_{\rm 
    pwd}), \quad t > 0,}
\label{eq:vda_analytical}
\end{equation}
\noindent{\added{as the classic model for heat conduction through a plane wall 
\citep{bergman2011fundamentals}. Here, the plane wall in $(0,s_{\rm pwd})$ 
represents the region of the powder-bed subject to relevant thermal effects of 
the printing process (i.e. with presence of significant thermal gradients). The 
wall is in contact with the surface of the part at $x = 0$ and has average 
length $s_{\rm pwd}$, referred to as the effective thermal thickness. A point 
in the powder bed that distances itself more than $s_{\rm pwd}$ with respect to 
the part is barely affected by the thermal gradient and undergoes little 
changes in temperature during the printing process, i.e. it mostly remains at 
$T_0$.}}

\deleted{According to this, the setting considers a 1D spring with material 
properties at powder state, subject to thermal contact at both ends.} 
\added{The initial condition is the same one as in 
Eq.~\eqref{eq:balanceofenergy}, i.e. $T(0) = T_0$. Besides, thermal contact 
holds at both surfaces of the $s_{\rm pwd}$-thick powder-bed wall:} At one 
\replaced{side ($x = 0$)}{end of the spring}, the powder-bed surface is in 
contact with the solid (part or plate) surface. As the powder-bed is a porous 
medium, it is reasonable to assume that the contacting surfaces do not match 
perfectly. In particular, the standard Fourier law does not hold. Hence, heat 
flux is computed as the product of an \ac{htc} $h_{\rm s/p}$ and the thermal 
gap 
$(T_{\rm solid}(t) - T_{\rm pwd}^{\rm s/p}(t))$ between both surfaces, i.e.
\begin{equation}
    q_{\rm s/p}(T,t) = h_{\rm s/p} (T_{\rm solid} - T_{\rm pwd}^{\rm s/p}), 
    \enspace \text{in} \enspace x = 0, \enspace t > 0.
\label{eq:thermal_contact_s}
\end{equation}

\deleted{The spring represents the region of the powder-bed subject to relevant 
thermal effects of the printing process (i.e. with presence of significant 
thermal gradients). It has average length $s_{\rm pwd}$, referred to as the 
effective thermal thickness. A point in the powder bed that distances itself 
more than $s_{\rm pwd}$ with respect to the part is barely affected by the 
thermal gradient and undergoes little changes in temperature during the 
printing process, i.e. it mostly remains at $T_0$.}

At the other \replaced{side ($x = s_{\rm pwd}$)}{end of the spring}, thermal 
powder-to-powder contact applies. The expression for the heat flux is analogous 
to Eq.~\eqref{eq:thermal_contact_s}; in this case, the \ac{htc} is $h_{\rm 
p/p}$ and the thermal gap can be taken as $(T_{\rm pwd}^{\rm p/p}(t) - T_0)$, 
in agreement with the discussion \replaced{above}{at the previous paragraph}. 
Hence,
\begin{equation}
    q_{\rm p/p}(T,t) = h_{\rm p/p} (T_{\rm pwd}^{\rm p/p} - T_0),
    \enspace \text{in} \enspace x = s_{\rm pwd}, \enspace t > 0.
\label{eq:thermal_contact_p}
\end{equation}

Assuming now a \ac{vda} \added{problem} attached to each integration point 
located on a face in $\partial \Omega_{\rm bed}^{\rm part} \cup \partial 
\Omega_{\rm bed}^{\rm base}$, $T_{\rm solid}\added{(t)}$ is determined from the 
solution of Eq.~\eqref{eq:balanceofenergy} \added{at any $t > 0$}, whereas the 
material properties of the powder, i.e. $\rho_{\rm pwd}$, $c_{\rm pwd}$ and 
$k_{\rm pwd}$, and the parameters $s_{\rm pwd}$, $h_{\rm s/p}$ and $h_{\rm 
p/p}$ are assumed to be known data. This leaves the \ac{vda} model as a simple 
well-posed 1D transient heat transfer problem. The first step of the \ac{vda} 
method is to apply a suitable discretization of this 1D problem. The resulting 
linear system is then modified, by adding $T_{\rm solid}$ as an additional 
unknown with the extra equation given by Eq.~\eqref{eq:thermal_contact_s}. 
Following this, static condensation allows one to recover $q_{\rm s/p}$, such 
that it no longer depends on $T_{\rm pwd}^{\rm s/p}$, only on $T_{\rm solid}$, 
the known data and the type of discretization. In this way, an equivalent 
expression of Eq.~\eqref{eq:thermal_contact_s} is obtained, that takes the same 
form as Eq.~\eqref{eq:bondary_conditions} and can be straightforwardly 
evaluated in the discrete form of Eq.~\eqref{eq:balanceofenergy}.

Let us illustrate the procedure considering (1) a discretization in space of 
\replaced{Eq.~\eqref{eq:vda_analytical}}{the spring} with a single linear 
Lagrangian finite element along the thickness $s_{\rm pwd}$; (2) a forward 
first order finite difference to approximate the time derivative; and (3) 
constant material properties. For this setting, the linear system of the 
\ac{vda} problem at the $n$-th time step ($n \geq 0$) 
is
\begin{equation}
	\begin{aligned}
		& \left[
			\begin{array}{cc}
				 h_{\rm s/p} + \frac{M_{11}}{\Delta t} + K_{11} & K_{12} \\ 
				 K_{21} & \frac{M_{22}}{\Delta t} + K_{22} + h_{\rm p/p}
			\end{array}
		\right]
		\left[ 
			\begin{array}{c}
				T_{\rm pwd}^{\rm{s/p},{n+1}} \\ 
				T_{\rm pwd}^{\rm{p/p},{n+1}}
			\end{array}
		\right] = \left[ 
			\begin{array}{c}
				\frac{M_{11}}{\Delta t} T_{\rm pwd}^{\rm{s/p},n} + h_{\rm s/p} 
				T_{\rm solid}^{n+1} \\ 
				\frac{M_{22}}{\Delta t} T_{\rm pwd}^{\rm{p/p},n}+ h_{\rm p/p} 
				T_0 \\ 
			\end{array}
		\right],
	\end{aligned}
\label{eq:vda_linear_system}
\end{equation}
\noindent{where $M_{ij}$ and $K_{ij}$ are the coefficients of the mass 
$\mathbf{M}$ and conductivity $\mathbf{K}$ matrices for the 1D linear 
lagrangian \ac{fe} of the \ac{vda} problem, with}
\begin{equation}
	\mathbf{M} = \rho_{\rm pwd} c_{\rm pwd} \frac{s_{\rm pwd}}{2} \left[
		\begin{array}{cc}
			 1 & 0 \\ 
			 0 & 1
		\end{array}
	\right]
	\quad \text{and} \quad
	\mathbf{K} = \frac{k_{\rm pwd}}{s_{\rm pwd}} \left[
		\begin{array}{cc}
			 1 & -1 \\ 
			 -1 & 1
		\end{array}
	\right].
\label{eq:vda_matrices}
\end{equation}

If $T_{\rm solid}^{n+1}$ is reinterpreted as an unknown, then, taking 
Eq.~\eqref{eq:thermal_contact_s} as an extra equation of the system, 
Eq.~\eqref{eq:vda_linear_system} is augmented to
\begin{equation}
	\begin{aligned}
	&\left[
		\begin{array}{ccc}
			h_{\rm s/p} & - h_{\rm s/p} & 0 \\
			- h_{\rm s/p} & h_{\rm s/p} + \frac{M_{11}}{\Delta t} + K_{11} & 
			K_{12} \\ 
			0 & K_{21} & \frac{M_{22}}{\Delta t} + K_{22} + h_{\rm p/p}
		\end{array}
	\right]
	\left[ 
		\begin{array}{c}
		    T_{\rm solid}^{n+1} \\
			T_{\rm pwd}^{\rm{s/p},{n+1}} \\ 
			T_{\rm pwd}^{\rm{p/p},{n+1}}
		\end{array}
	\right] = \left[ 
		\begin{array}{c}
		    q_{\rm s/p}^{n+1} \\
			\frac{M_{11}}{\Delta t} T_{\rm pwd}^{\rm{s/p},n} \\ 
			\frac{M_{22}}{\Delta t} T_{\rm pwd}^{\rm{p/p},n}+ h_{\rm p/p} 
			T_0 \\ 
		\end{array}
	\right],
	\end{aligned}
\label{eq:vda_linear_system_extended}
\end{equation}

Identifying now the blocks
\begin{align}
	& \mathbf{A}_{12} = [ \ -h_{\rm s/p} \ 0 \ ], \\ & \mathbf{A}_{21} = 
	\mathbf{A}_{12}^{tr}, \\ & \mathbf{A}_{22} = \left[
		\begin{array}{cc}
		 h_{\rm s/p} + \frac{M_{11}}{\Delta t} + K_{11} & K_{12} \\ 
		 K_{21} & \frac{M_{22}}{\Delta t} + K_{22} + h_{\rm p/p}
		\end{array}
	\right] \ \text{and} \\ & \mathbf{b}_{2} = \left[ 
		\begin{array}{c}
			\frac{M_{11}}{\Delta t} T_{\rm pwd}^{\rm{s/p},n} \\ 
			\frac{M_{22}}{\Delta t} T_{\rm pwd}^{\rm{p/p},n}+ h_{\rm p/p} 
			T_0
		\end{array}
	\right],
\label{eq:vda_blocks}
\end{align}
\noindent{Eq.~\eqref{eq:vda_linear_system_extended} can be rewritten as} 
\begin{equation}
	\left[
		\begin{array}{cc}
			h_{\rm s/p} & \mathbf{A}_{12} \\
			\mathbf{A}_{21} & \mathbf{A}_{22} \\
		\end{array}
	\right]
	\left[ 
		\begin{array}{c}
		    T_{\rm solid}^{n+1} \\
			\mathbf{T}_{\rm pwd}^{n+1}
		\end{array}
	\right] = \left[ 
		\begin{array}{c}
		    q_{\rm s/p}^{n+1} \\
			\mathbf{b}_2 \\ 
		\end{array}
	\right].
\label{eq:vda_linear_system_blocks}
\end{equation}

Following this, application of static condensation to 
Eq.~\eqref{eq:vda_linear_system_blocks} to remove the unknowns of 
$\mathbf{T}_{\rm pwd}^{n+1}$, leads to the relation
\begin{equation}
	\begin{aligned}
		&\left[ h_{\rm s/p} - \mathbf{A}_{12} (\mathbf{A}_{22})^{-1} 
		\mathbf{A}_{21} \right] T_{\rm solid}^{n+1} \\ &\quad = q_{\rm s/p}^{n+1} - 
		\mathbf{A}_{12} (\mathbf{A}_{22})^{-1} \mathbf{b}_{2}.
	\end{aligned}
\label{eq:vda_static_condensation}
\end{equation}
From this point, it suffices to compare Eq.~\eqref{eq:vda_static_condensation} 
with Eq.~\eqref{eq:bondary_conditions} to see that
\begin{equation}
	\begin{aligned}
   	h_{\rm loss} &= h_{\rm s/p} - \mathbf{A}_{12} (\mathbf{A}_{22})^{-1} 
   	\mathbf{A}_{21} \\
		T_{\rm loss} &= - \left. \mathbf{A}_{12} (\mathbf{A}_{22})^{-1} 
		\mathbf{b}_{2} \ \right/ h_{\rm loss} 
	\end{aligned}
\label{eq:vda_equivalent}
\end{equation}

The \ac{vda} method can be applied verbatim (with enhanced accuracy), if 
several and/or higher order \acp{fe} are considered; some examples are listed 
in Tab.~\ref{tab:vda_equations}. The reason is that the discretization 
leads to the same block structure of Eq.~\eqref{eq:vda_linear_system_blocks}.  
Likewise, the part-plate thermal contact in the \ac{p} model can be analogously 
constructed; in this case, the material properties are those of the plate.

\newcommand{\hs}{h_{\rm s/p}}
\newcommand{\he}{h_{\rm p/p}}
\newcommand{\cc}{m}
\newcommand{\kk}{\hat{k}}
\newcommand{\tsn}{T_{\rm pwd}^{\rm{s/p},n}}
\newcommand{\tson}{T_{\rm solid}^n}
\newcommand{\tmn}{T_{\rm pwd}^{\rm{m},n}}
\newcommand{\ten}{T_{\rm pwd}^{\rm{p/p},n}}
\newcommand{\te}{T_0}

\begin{table}[!h]
    \centering
    \begin{tabular}{ | m{0.96\textwidth} | }
        \hline
        One linear FE along the VDA \added{(1E-Q1)} \\ \hline
				\[
					\begin{aligned}
				   	h_{\rm loss} &= \frac{ 4 \hs ( \he + \cc ) \kk + 2 \cc  
				   	\hs \he + \cc^2 \hs }{ 4 ( \hs + \he + \cc ) \kk + ( 4 \hs + 
				   	2 \cc ) \he + 2 \cc \hs + \cc^2 } \\
						T_{\rm loss} &= \frac{ 4 \hs \he \kk \te + 2 \cc \hs \kk \ten + (2  
						\kk + 2 
						\he + \cc ) \cc \hs \tsn }{ 4 ( \hs + \he + \cc ) \kk + 4 ( \hs + 
						\frac{\cc}{2} ) \he + 2 \cc \hs + \cc^2 }
					\end{aligned}
				\] \\ \hline
        Two linear FE along the VDA \added{(2E-Q1)} \\ \hline
        \scriptsize{
				\[
					\begin{aligned}
				   	h_{\rm loss} &= \frac{ 128 ( \he + \cc ) \hs \kk^2 + ( 64 \he + 24 
				   	\cc ) 
				   	\cc \hs \kk + 4 \cc^2 \hs \he + \cc^3 \hs }{ 128 ( \hs + \he + \cc 
				   	) \kk^2 
				   	+ ( ( 128 \hs + 64 \cc ) \he + 64 \cc \hs + 24 \cc^2 ) \kk + ( 16 
				   	\cc \hs + 
				   	4 \cc^2 ) \he + 4 \cc^2 \hs + \cc^3 } \\
						T_{\rm loss} &= \frac{ 128 \hs \he \kk^2 \te + 32 \cc \hs \kk^2 
						\ten + ( 64 
						\cc \hs \kk^2 + ( 32 \cc \hs \he + 8 \cc^2 \hs ) \kk ) \tmn }{ 128 
						( \hs + 
						\he + \cc ) \kk^2 + ( ( 128 \hs + 64 \cc ) \he + 64 \cc \hs + 24 
						\cc^2 ) 
						\kk + ( 16 \cc \hs + 4 \cc^2 ) \he + 4 \cc^2 \hs + \cc^3 } \\ &+ 
						\frac{ (32 
						\cc \hs \kk^2 + (32 \cc \hs \he + 16 \cc^2 \hs ) \kk + 4 \cc^2 \hs 
						\he + 
						\cc^3 \hs ) \tsn }{ 128 ( \hs + \he + \cc ) \kk^2 + ( ( 128 \hs + 
						64 \cc ) 
						\he + 64 \cc \hs + 24 \cc^2 ) \kk + ( 16 \cc \hs + 4 \cc^2 ) \he + 
						4 \cc^2 
						\hs + \cc^3 }
					\end{aligned}
				\]} \\ \hline
        One quadratic FE along the VDA \added{(1E-Q2)} \\ \hline
        \scriptsize{
				\[
					\begin{aligned}
				   	h_{\rm loss} &= \frac{ 288 ( \he + \cc ) \hs \kk^2 + ( 132 \he + 36 
				   	\cc ) 
				   	\cc \hs \kk + 6 \cc^2 \hs \he + \cc^3 \hs }{ 288 ( \hs + \he + \cc 
				   	) \kk^2 
				   	+ ( ( 288 \hs + 132 \cc ) \he + 132 \cc \hs + 36 \cc^2 ) \kk + ( 36 
				   	\cc \hs 
				   	+ 6 \cc^2 ) \he + 6 \cc^2 \hs + \cc^3 } \\
						T_{\rm loss} &= \frac{ ( 288 \kk^2 - 12 \cc \kk ) \hs \he \te + ( 
						48 \cc 
						\kk^2 - 2 \cc^2 \hs ) \kk \ten + ( 192 \cc \hs \kk^2 + ( 96 \cc \hs 
						\he + 
						16 \cc^2 \hs ) \kk ) \tmn }{ 288 ( \hs + 		\he + \cc ) \kk^2 + ( ( 
						288 
						\hs + 132 \cc ) \he + 132 \cc \hs + 36 \cc^2 ) \kk + ( 36 \cc \hs + 
						6 \cc^2 
						) \he + 6 \cc^2 \hs + \cc^3 } \\ &+ \frac{ (48 \cc \hs \kk^2 + (48 
						\cc \hs 
						\he + 22 \cc^2 \hs ) \kk + 6 \cc^2 \hs \he + \cc^3 \hs ) \tsn }{ 
						288 ( \hs 
						+ \he + \cc ) \kk^2 + ( ( 288 \hs + 132 \cc ) \he + 132 \cc \hs + 
						36 \cc^2 
						) \kk + ( 36 \cc \hs + 6 \cc^2 ) \he + 6 \cc^2 \hs + \cc^3 }
					\end{aligned}
				\]} \\ \hline
        One linear \ac{fe} along the \ac{vda}, $T_{\rm pwd}^{\rm s/p} 
        \approx T_{\rm solid}$ and $T_{\rm pwd}^{\rm p/p} \approx T_0$ 
        \added{(1E-Q1-D)} \\ \hline
				\[
					\begin{aligned}
				   	h_{\rm loss} &= \frac{1}{2} \cc + \kk \\
						h_{\rm loss} T_{\rm loss} &= \frac{1}{2} \cc \tson + \kk \te
					\end{aligned}
				\] \\ \hline
    \end{tabular}
    \vspace{0.2cm}
    \caption{Expressions of $h_{\rm loss}$ and $T_{\rm loss}$ in terms of the 
    \ac{vda} parameters for different types of discretizations. Forward first 
    order finite difference in time and constant material properties are 
    assumed. $m \vcentcolon= \rho_{\rm pwd} c_{\rm pwd} s_{\rm pwd} / \Delta 
    t$, $\hat{k} \vcentcolon= k_{\rm pwd} / s_{\rm pwd}$ and $T_{\rm 
    pwd}^{\rm{m},n}$ is the temperature in the middle of the \ac{vda} at the 
    previous time step.}
    \label{tab:vda_equations}
\end{table}

But the main advantage of the \ac{vda} concerns computational efficiency. As 
the method ends up extracting a parametrized closed-form expression to 
evaluate the equivalent boundary condition, there is no need to 
solve a linear system at each integration point. The additional cost with 
respect to the \ac{htc} model merely consists in extra storage of the 
temperature values of the \ac{vda} at previous time steps (depending on the 
time integration scheme) and extra floating point operations to evaluate 
Eq.~\eqref{eq:vda_equivalent}.

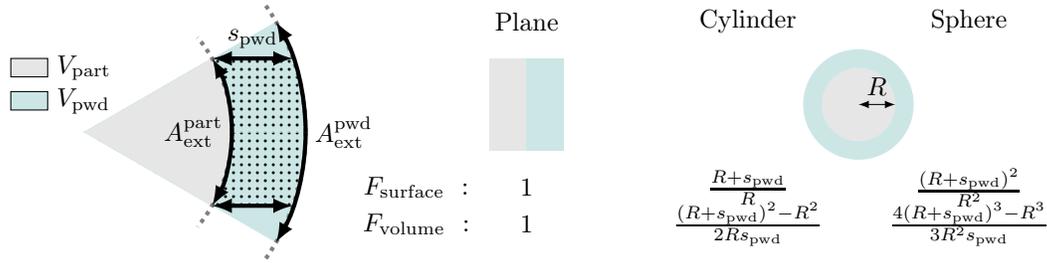
\begin{figure}[!h]
  \centering
  \resizebox{0.9\textwidth}{!}{%


\begin{tikzpicture}
	
	\filldraw[teal!20!white] (3,0.5) -- (6,0.5) arc[radius=3cm,start angle=0,end 
	angle=30];
	\filldraw[teal!20!white] (3,0.5) -- (6,0.5) arc[radius=3cm,start angle=0,end 
	angle=-30];
	\filldraw[teal!20!white, pattern=dots] (3,0.5) -- (6,0.5) arc[radius=3cm,start angle=0,end 
	angle=19.47] -- (4.732,1.5) -- cycle;
	\filldraw[teal!20!white, pattern=dots] (3,0.5) -- (6,0.5) arc[radius=3cm,start angle=0,end 
	angle=-19.47] -- (4.732,-0.5) -- cycle;
	\filldraw[gray!20!white] (3,0.5) -- (5,0.5) arc[radius=2cm,start angle=0,end 
	angle=30];
	\filldraw[gray!20!white] (3,0.5) -- (5,0.5) arc[radius=2cm,start angle=0,end 
	angle=-30];
	
	\node[anchor=east] at (5.0,0.5) {$A_{\rm ext}^{\rm part}$};
	\node[anchor=west] at (6.0,0.5) {$A_{\rm ext}^{\rm pwd}$};
	
	\draw[gray, ultra thick, dotted] (5,0.5) arc[radius=2cm,start angle=0,end 
	angle=40];
	\draw[gray, ultra thick, dotted] (5,0.5) arc[radius=2cm,start angle=0,end 
	angle=-40];
	\draw[black, ultra thick, -latex] (5,0.5) arc[radius=2cm,start angle=0,end 
	angle=30];
	\draw[black, ultra thick, -latex] (5,0.5) arc[radius=2cm,start angle=0,end 
	angle=-30];
	
	\draw[gray, ultra thick, dotted] (6,0.5) arc[radius=3cm,start angle=0,end 
	angle=35];
	\draw[gray, ultra thick, dotted] (6,0.5) arc[radius=3cm,start angle=0,end 
	angle=-35];
	\draw[black, ultra thick, -latex] (6,0.5) arc[radius=3cm,start angle=0,end 
	angle=30];
	\draw[black, ultra thick, -latex] (6,0.5) arc[radius=3cm,start angle=0,end 
	angle=-30];
	
	\draw[black, ultra thick,latex-latex] (4.732,1.5) -- node[anchor=south] {$s_{\rm pwd}$} (5.828,1.5);
	\draw[black, ultra thick,latex-latex] (4.732,-0.5) -- (5.828,-0.5);
	
	\filldraw[fill=gray!20!white] (2.0,1.5) -- (2.0,1.25) -- (2.5,1.25) -- node[anchor=west] {$V_{\rm part}$} (2.5,1.5) -- cycle;
	\filldraw[fill=teal!20!white] (2.0,1.05) -- (2.0,0.80) -- (2.5,0.80) -- node[anchor=west] {$V_{\rm pwd}$} (2.5,1.05) -- cycle;
	
	\node at (9.0,2.0) {Plane};
	\node at (12.0,2.0) {Cylinder};
	\node at (15.0,2.0) {Sphere};
	
	\node at (7.5,-0.25) {$F_{\rm surface}$ \ :};
	\node at (7.5,-0.75) {$F_{\rm volume}$ \ :};
	
	\fill[gray!20!white] (8.5,0.25) rectangle (9.5,1.5);
	\fill[teal!20!white] (9.0,0.25) rectangle (9.5,1.5);
	
	\node at (9.0,-0.25) {1};
	\node at (9.0,-0.75) {1};
	
	\fill[teal!20!white] (13.5,0.875) circle [radius=0.75];
	\fill[gray!20!white] (13.5,0.875) circle [radius=0.5];	
	
	\draw[latex-latex] (13.5,0.875) -- node[anchor=south] {$R$} (14.0,0.875);
	
	\node at (12.0,-0.25) {$\frac{R+s_{\rm pwd}}{R}$};
	\node at (12.0,-0.75) {$\frac{(R+s_{\rm pwd})^2-R^2}{2 R s_{\rm pwd}}$};
	
	\node at (15.0,-0.25) {$\frac{(R+s_{\rm pwd})^2}{R^2}$};
	\node at (15.0,-0.75) {$\frac{4(R+s_{\rm pwd})^3-R^3}{3 R^2 s_{\rm pwd}}$};
	
\end{tikzpicture}%
  }%
  \caption{Geometrical correction factors to account for non-unidimensional 
  heat transfer.}
  \label{fig:vda_geometrical_factors}
\end{figure}

The main limitation of the method is that the contour of the part-powder 
interface is often a smooth 2D shape and, as a result, heat transfer 
across the powder-bed is not unidimensional and orthogonal to the interface. 
{Depending on the curvature of the geometry, heat diffusion through 
directions other than the normal to the discrete boundary may be relevant and 
the \ac{vda} risks of underestimating the amount of heat loss.} For simple 
shapes (e.g. cylinder or sphere) the local curvature of $\partial \Omega_{\rm 
bed}^{\rm part} \cup \partial \Omega_{\rm bed}^{\rm base}$ can be taken into 
account by considering a modified 1D heat transfer problem with
\begin{equation}
	\begin{aligned}
    \hat{\rho}_{\rm pwd} \hat{c}_{\rm pwd} &= \rho_{\rm pwd} c_{\rm pwd} F_{\rm 
    volume}, \\ \hat{k}_{\rm pwd} &= k_{\rm pwd} F_{\rm volume}, \text{ and } \\
    \hat{h}_{\rm loss} &= h_{\rm loss} F_{\rm surface},
  \end{aligned}
\label{eq:vda_modified}
\end{equation}
with the geometrical correction factors $F_{\rm volume}$ and $F_{\rm 
surface}$ defined as
\begin{equation}
     F_{\rm volume} = \frac{V_{\rm pwd}}{A_{\rm ext}^{\rm part} s_{\rm pwd}} 
     \quad \text{and} \quad F_{\rm surface} = \frac{A_{\rm ext}^{\rm 
     pwd}}{A_{\rm ext}^{\rm part}},
\label{eq:vda_factors}
\end{equation}
where, as shown in Fig.~\ref{fig:vda_geometrical_factors}, $V_{\rm pwd}$ is 
the volume of the fraction of powder bed modelled by the \ac{vda} model and 
$A_{\rm ext}^{\rm pwd}$ and $A_{\rm ext}^{\rm part}$ are the areas of the 
external surfaces of the \ac{vda} and the part. {These corrections are 
intended to compensate for the nonorthogonal heat loss neglected by the 
standard approach.}

Another drawback is that the \ac{vda}, in spite of adding physics into 
the computation of the boundary condition, still needs to estimate some 
unknown quantities, such as $h_{\rm s/p}$, $h_{\rm p/p}$ or $s_{\rm pwd}$. 
Assuming that $T_{\rm pwd}^{\rm s/p} \approx T_{\rm solid}$ and $T_{\rm 
pwd}^{\rm p/p} \approx T_0$, one can get rid of estimating the \acp{htc}, 
because the boundary conditions at both ends of the \ac{vda} problem
become of Dirichlet type. A relation like Eq.~\eqref{eq:bondary_conditions} 
is then obtained following exactly the same procedure outlined above. 
Additionally, the determination of $s_{\rm pwd}$ is object of discussion in 
Sect.~\ref{sec:verification}.

\section{Numerical experiments}
\label{sec:numerical}

\subsection{Verification of the VDA against the HF model}
\label{sec:verification}

The purpose of the first numerical experiment is to verify the new \ac{vda} 
formulation and demonstrate its capabilities and benefits in the thermal 
simulation of a \ac{slm} or \ac{dmls} process. Three model variants 
(cf. Sect.~\ref{sec:variants}) are considered: (1) \ac{hf}, (2) 
\ac{vda}-\ac{pp} and (3) \ac{vda}-\ac{p}. The first one is taken as the 
numerical reference for the other two, due to its higher physical 
representativeness.

\begin{figure}[!h]
  \centering
  \begin{subfigure}[t]{0.45\textwidth}
    \includegraphics[width=\textwidth]{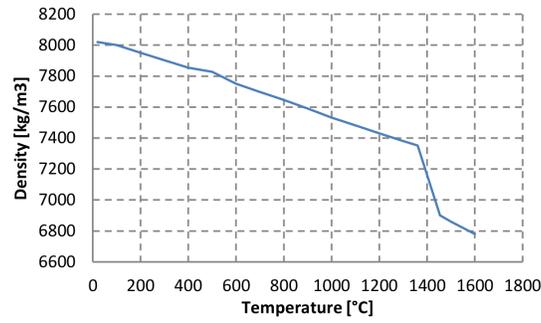}
    \caption{Density.}
  \end{subfigure}
  \\
  \begin{subfigure}[t]{0.45\textwidth}
    \includegraphics[width=\textwidth]{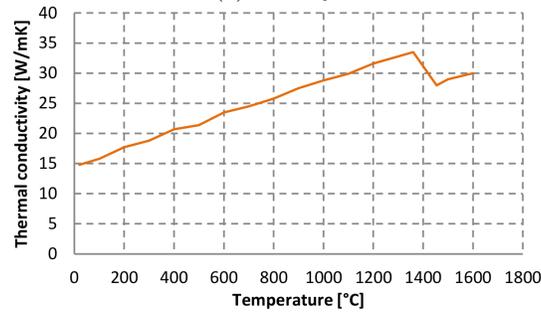}
    \caption{Thermal conductivity.}
  \end{subfigure}
  \\
  \begin{subfigure}[t]{0.45\textwidth}
    \includegraphics[width=\textwidth]{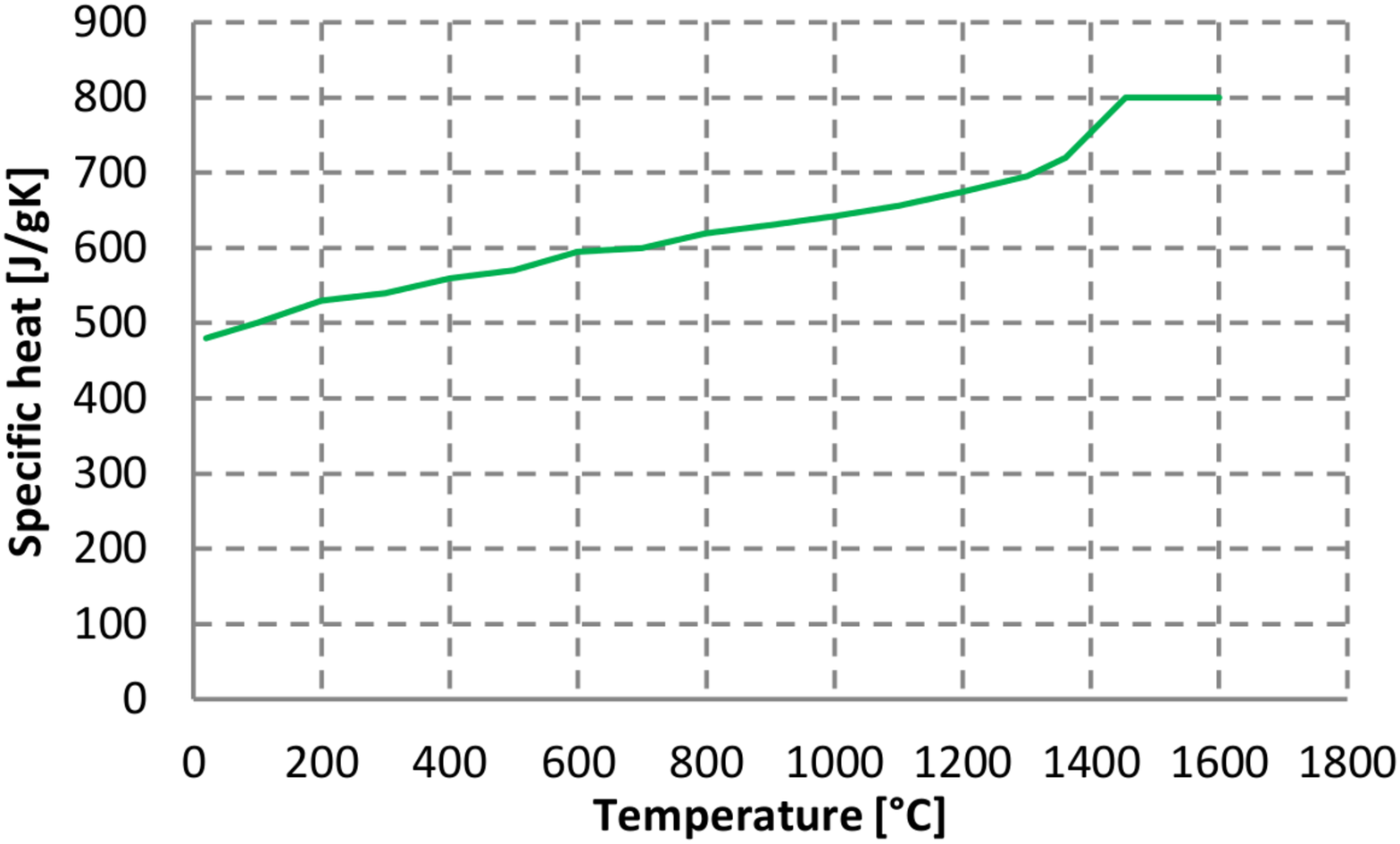}
    \caption{Specific heat.}
  \end{subfigure}
  \caption{Stainless Steel 304L thermal bulk material properties 
  \citep{mills2002recommended}.}
  \label{fig:SS304L}
\end{figure}

The example is designed to be simple, but suitable enough to compare the 
accuracy and efficiency of models (2) and (3) with respect to model (1). 
According to this, the object of simulation is the printing of a 10x10x10 
$[\mathrm{mm}^3]$ cube on top of a 110x110x20 $[\mathrm{mm}^3]$ metal 
substrate. Different materials are selected for the cubic sample and the 
build plate: while the former is made of Maraging steel M300, the latter is 
composed of Stainless Steel (SS) 304L. Bulk temperature-dependent thermal 
properties of SS 304L~\citep{mills2002recommended} are represented in 
Fig.~\ref{fig:SS304L}. On the other hand, Tab.~\ref{tab:M300} prescribes 
constant density and specific heat values for M300, due to lack of 
temperature-dependent data, and the M300 powder is assumed to have 54 \% of 
relative density.

\begin{table}[!h]
    \centering
    \begin{tabular}{ | c | c | c | c | }
        \hline
        \textbf{Temperature} $\mathbf{[C]}$ & \textbf{Density} 
        $\mathbf{[kg/m^3]}$ & \textbf{Specific heat} $\mathbf{[J/gC]}$ & 
        \textbf{Conductivity $\mathbf{[W/mC]}$} \\ \hline
        20.0 & \multirow{4}{*}{8,100} & \multirow{4}{*}{500} & 14.2 \\ 
        \cline{1-1} \cline{4-4}
        600.0 & & & 21.0 \\ \cline{1-1} \cline{4-4}
        1,300.0 & & & 28.6 \\ \cline{1-1} \cline{4-4}
        1,600.0 & & & 28.6 \\ \hline
    \end{tabular}
    \vspace{0.2cm}
    \caption{Maraging steel M300 thermal bulk material properties
    \citep{renishaw}.}
    \label{tab:M300}
\end{table}

The printing process considers a layer thickness of 30 $[\upmu \mathrm{m}]$. 
Therefore, a total number of 333 layers are deposited to build the sample. 
The values of the remaining process parameters are detailed in 
Tab.~\ref{tab:process_verif}. Note that there is no information regarding the 
scanning path, because the simulation considers a layer-by-layer deposition 
sequence. This means that the printing of a complete layer is simulated in a 
single time step. As a result, the simulation follows an alternating sequence 
in time consisting of:
\begin{enumerate} 
	\item \textbf{Printing step:} A new layer is activated, i.e. added 
	into the computational domain, and the problem is solved applying the 
	energy input necessary to fuse the powder of the whole layer. The time 
	increment is given by the \emph{scanning time}.
	\item \textbf{Cooling step:} The problem is solved without heat 
	application to account for the time lowering the plate, recoat time 
	and laser relocation. Here, the time step is the \emph{recoating time}. %
\end{enumerate}

\begin{table}[!h]
    \centering
    \begin{tabular}{ | c | c c | }
        \hline
        Power input & $375$ & $[\mathrm{W}]$ \\ \hline
        Scanning time & $2$ & $[\mathrm{s}]$ \\ \hline
        Recoating time & $10$ & $[\mathrm{s}]$ \\ \hline
        Absorption coefficient & $0.64$ & - \\ \hline
        Layer thickness & $30$ & $[\upmu \mathrm{m}]$ \\ \hline
        Plate temperature & $20$ & $[\mathrm{C}]$ \\ \hline
    \end{tabular}
    \vspace{0.2cm}
    \caption{Process parameters for the example in 
    Sect.~\ref{sec:verification}.} 
    \label{tab:process_verif}
\end{table}

\begin{figure}[!h]
  \centering
  \begin{subfigure}[t]{0.3\textwidth}
    \includegraphics[width=\textwidth]{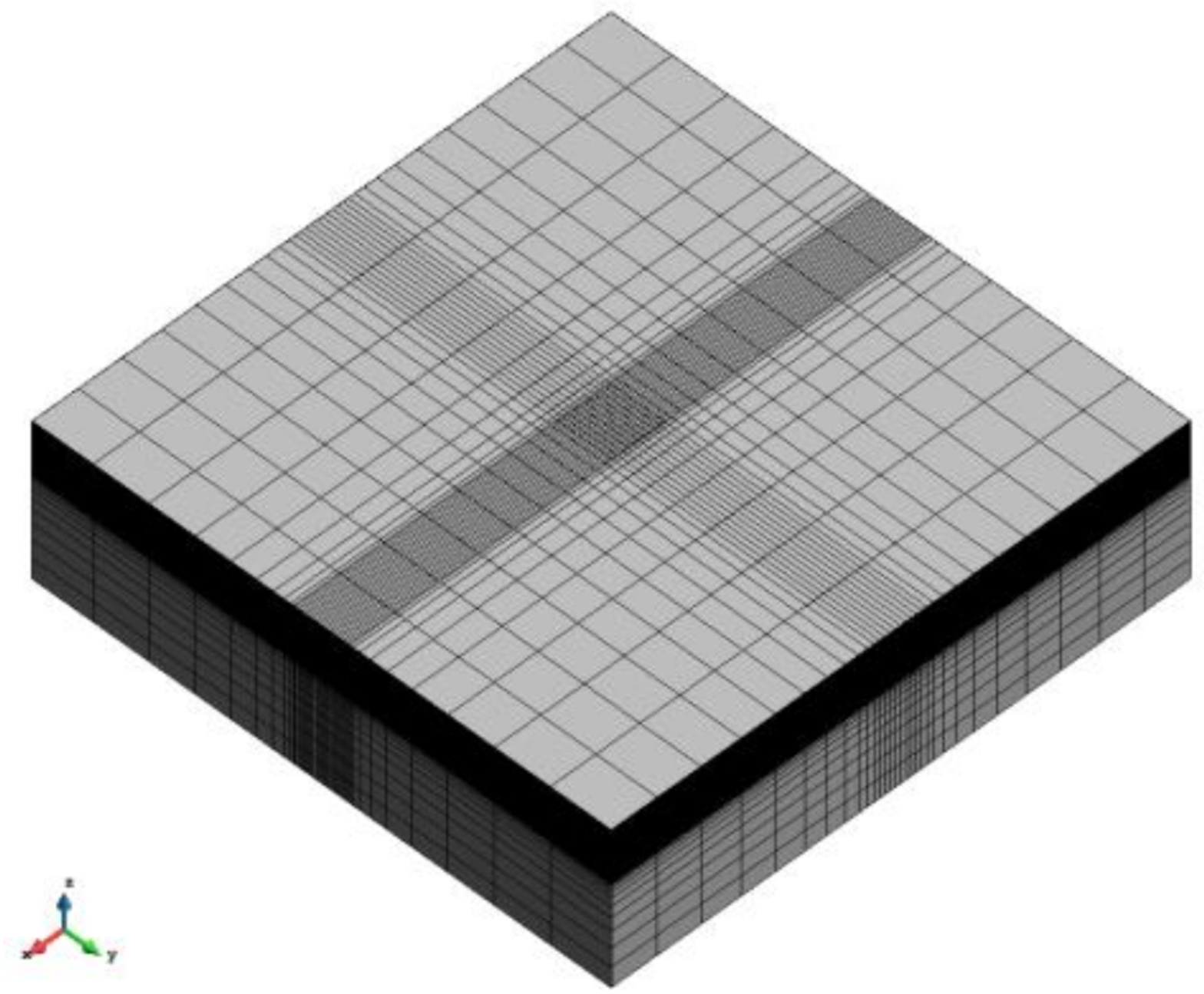}
    \caption{\ac{hf} model.}
  \end{subfigure} \quad 
  \begin{subfigure}[t]{0.3\textwidth}
    \includegraphics[width=\textwidth]{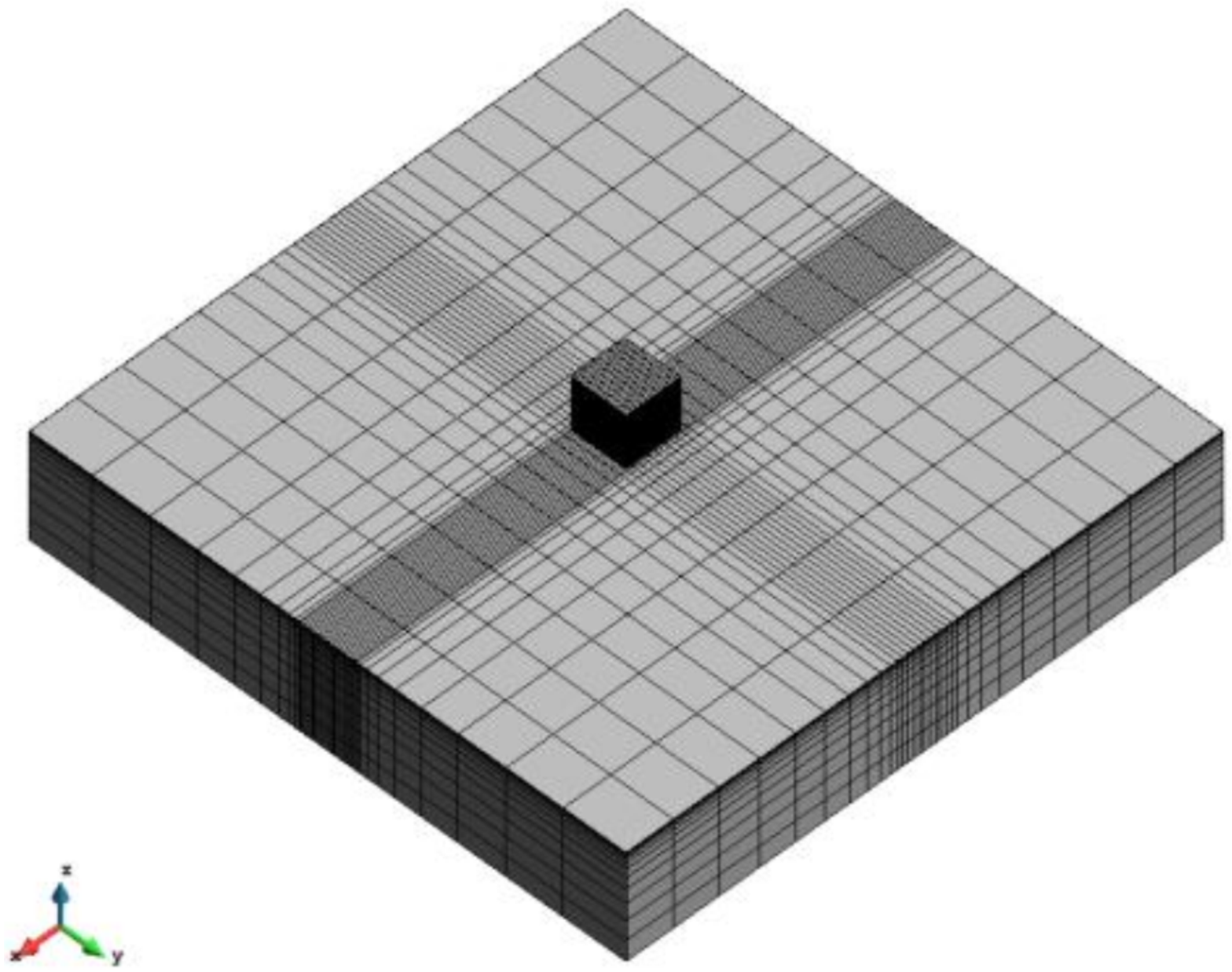}
    \caption{\ac{vda}-\ac{pp} model.}
  \end{subfigure} \quad
  \begin{subfigure}[t]{0.2\textwidth}
    \includegraphics[width=\textwidth]{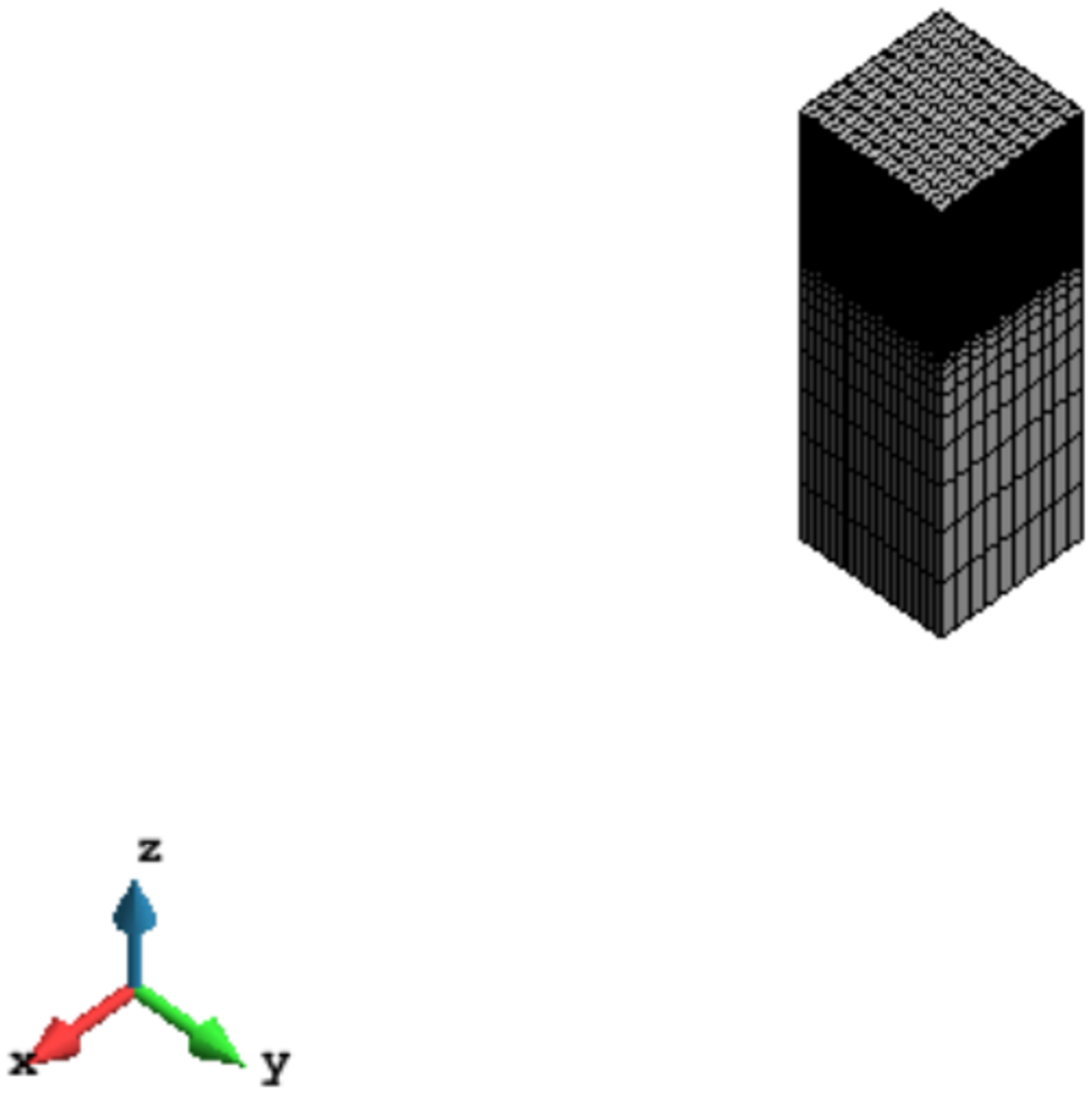}
    \caption{\ac{vda}-\ac{p} model.}
  \end{subfigure}
  \caption{FE meshes of the three models tested in the example of 
  Sect.~\ref{sec:verification}.}
  \label{fig:verification_models}
\end{figure}

Fig.~\ref{fig:verification_models} shows the FE discretizations of the three 
tested models. All three cases consider structured meshes of varying size to 
adequately capture the physics of the process. In particular, a finer 
layer-conforming mesh is prescribed for the fabricated part, whereas a 
coarser mesh is used away from the printed part. As observed in 
Tab.~\ref{tab:meshes_time}, the mesh size decreases one order of magnitude 
from the \ac{hf} model to the \ac{vda}-\ac{pp} and \ac{vda}-\ac{p} ones.

\begin{table}[!h]
    \centering
    \begin{tabular}{ | c | c | c | }
        \hline
        \textbf{Model} & \textbf{Mesh size [DoFs]} & \textbf{Runtime 
        [$\mathbf{h}$]-[\%]} \\ \hline
        \ac{hf} & 274.5k & 17.6 - 100 \\ \hline
        \ac{vda}-\ac{pp} & 66.5k & 1.3 - 7.4 \\ \hline
        \ac{vda}-\ac{p} & 49.9k & 0.8 - 4.5 \\ \hline
    \end{tabular}
    \vspace{0.2cm}
    \caption{Example of Sect.~\ref{sec:verification}. Mesh sizes and 
    simulation times of the reduced models are one order of magnitude down 
    the reference model.}
    \label{tab:meshes_time}
\end{table}

Concerning the boundary conditions, the top surfaces of the three models 
are subject to heat transfer through the surrounding air with an \ac{htc} of
10 [W/m\textsuperscript{2}C] and air temperature of 20 [C]. On the other 
hand, the same \ac{htc} and environment temperature are assigned at the 
lateral and bottom walls of the build chamber. 

The last ingredient is to characterize the BCs at the solid-powder 
and part-plate contacting surfaces, by establishing the \ac{vda} parameters 
for the \ac{vda} solid-powder (virtual powder) and the \ac{vda} part-plate 
(virtual plate). Recall that only the first applies to the \ac{vda}-\ac{pp} 
model, whereas both apply to the \ac{vda}-\ac{p} one. 

Both \acp{vda} adopt the full Dirichlet variant \added{ \added{(cf. (1E-Q1-D) 
in Tab.~\ref{tab:vda_equations})}} described at the end of 
Sect.~\ref{sec:virtual} with a single linear Lagrangian \ac{fe} and without 
geometrical correction factors\added{, for the sake of reducing the number of 
unknown parameters}. \replaced{In particular,}{This means that} the only 
\replaced{ones}{parameters} that need estimation are the virtual domain 
thicknesses $s_{\rm pwd}$ and $s_{\rm plate}$ and the virtual domain 
environment temperature $T_0$.

\begin{figure}[!h]
  \centering
  \includegraphics[width=0.8\textwidth]{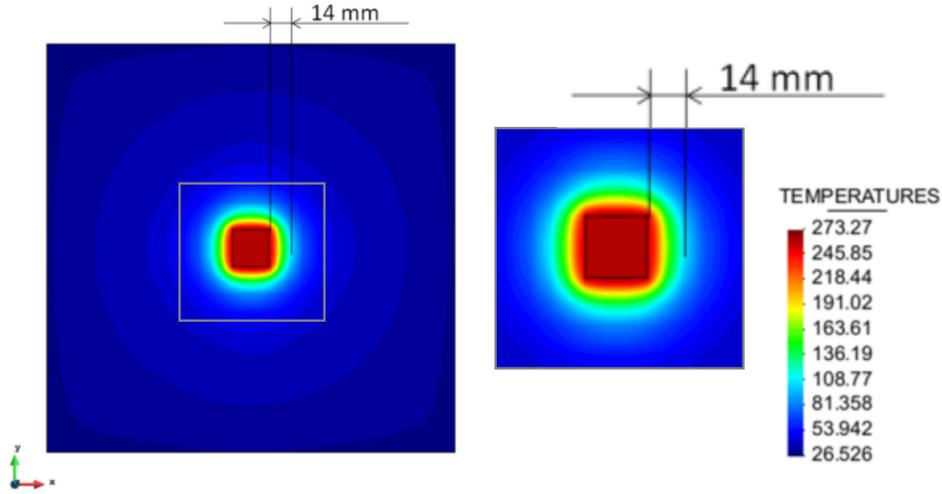}
  \caption{Contour plot of temperatures of the \ac{hf} model (XY view + close 
  up), showing the extent of the region concentrating thermal gradients.}
  \label{fig:ContourPlotVerif}
\end{figure}

The strategy for such estimation is based on simple calibration with respect 
to the \ac{hf} model. Let us explain the procedure step-by-step to model the 
virtual powder (analogous for the virtual plate). First, the temperature 
distribution of the \ac{hf} model is analysed to identify the region 
concentrating the strongest gradients. As shown in 
Fig.~\ref{fig:ContourPlotVerif} for the virtual 
powder, at about 14 $[\mathrm{mm}]$ away from the part, the temperature drops 
to 90 [C]. Further away, the thermal gradients are apparently smoother. 
Therefore, $T_0$ is set to 90 [C] and $s_{\rm pwd}$ is initially 
approximated by 14 $[\mathrm{mm}]$. Calibration with respect to the thermal 
history of the \ac{hf} model at a selected point follows to correct $s_{\rm 
pwd}$ to the final value. \ac{vda} parameter values obtained with this 
approach are listed in Tab.~\ref{tab:vda_params}.

\begin{table}[!h]
    \centering
    \begin{tabular}{ | c | c | c | }
        \hline
        \textbf{Domain} & \textbf{\ac{vda} thickness} $\mathbf{[mm]}$ & 
        $\mathbf{T_0 \ [C]}$ \\ \hline
        Powder & 10 & 90 \\ \hline
        Plate & 11 & 20 \\ \hline
    \end{tabular}
    \vspace{0.2cm}
    \caption{Example of Sect.~\ref{sec:verification}. \ac{vda} parameters
    after calibration with respect to the reference model.}
    \label{tab:vda_params}
\end{table}

\begin{figure}[!h]
  \centering
  \begin{subfigure}[t]{\textwidth}
    \begin{center}
	    \resizebox{\textwidth}{!}{
	    \input{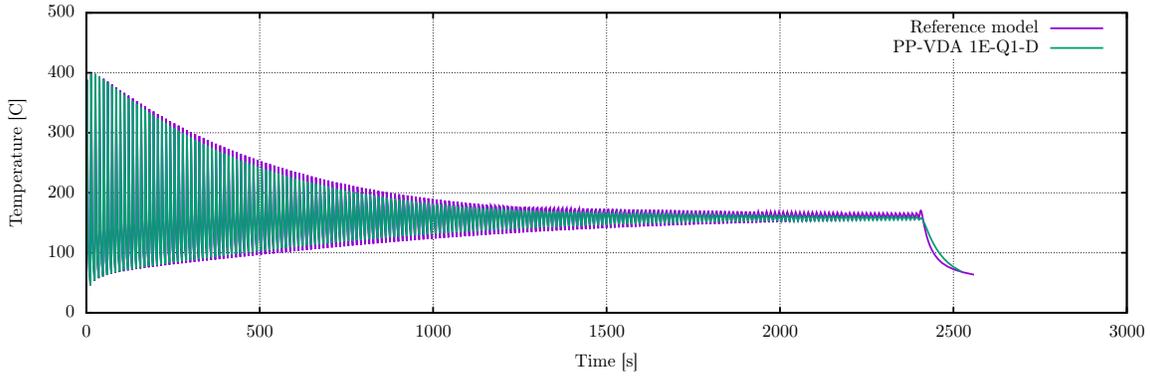}
	    }
	    \caption{\added{\ac{vda}-\ac{pp} model.}}
    \end{center}
  \end{subfigure}
  \\
  \begin{subfigure}[t]{\textwidth}
    \begin{center}
	    \resizebox{\textwidth}{!}{
	    \input{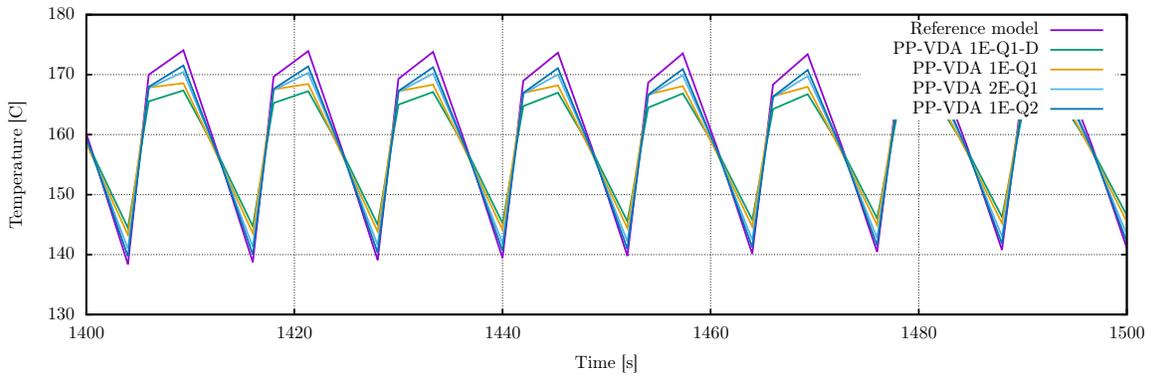}
	    }
	    \caption{\added{Close up of \ac{vda}-\ac{pp} model halfway through the 
	    simulation. Comparison against other \ac{vda} variants. Variant names in 
	    the key are the ones given in Tab.~\ref{tab:vda_equations}.}}
	    \label{fig:verification_pp_comp}
    \end{center}
  \end{subfigure}
  \caption{\added{Example of Sect.~\ref{sec:verification}. Comparison of the 
  reduced \ac{vda}-\ac{pp} model against the \ac{hf} model.}}
  \label{fig:verification_pp}
\end{figure}

The numerical experiments for this example are supported by the in-house 
research software COupled MEchanical and Thermal (COMET)
\citep{cervera2002comet} and GiD~\citep{GiDUserManual,gidhome} as a pre- 
and postprocessing software. Figs.~\ref{fig:verification_pp} 
and~\ref{fig:verification_p} compare the temperature evolutions of the reduced 
models against the reference \ac{hf} model, at a point located in the middle of 
the bottom surface of the part. As observed, both \ac{vda}-\ac{pp} and 
\ac{vda}-\ac{p} are  capable of reproducing the thermal response of the \ac{hf} 
model with errors bounded by 10 \% and 20 \% \added{with the Dirichlet 
\ac{vda} variant}. The \ac{vda}-\ac{p} is clearly a little less accurate than 
the former, as expected. However, as seen in Tab.~\ref{tab:meshes_time}, the 
computational running times of the \ac{vda} models are one order of magnitude 
less than the \ac{hf} model. This showcases the ability of the new formulation 
to approximate the response of a high-fidelity model, but with significantly 
increased efficiency, and it opens the path for larger scale and 
experimentally-based simulations, such as the one in the next section.

\begin{figure}[!h]
  \centering
  \begin{subfigure}[t]{\textwidth}
    \begin{center}
	    \resizebox{\textwidth}{!}{
	    \input{GAB-P.tex}
	    }
	    \caption{\added{\ac{vda}-\ac{p} model.}}
    \end{center}
  \end{subfigure}
  \\
  \begin{subfigure}[t]{\textwidth}
    \begin{center}
	    \resizebox{\textwidth}{!}{
	    \input{GAB-P-closeUP.tex}
	    }
	    \caption{\added{Close up of \ac{vda}-\ac{p} model halfway through the 
	    simulation.}}
    \end{center}
  \end{subfigure}
  \caption{\added{Example of Sect.~\ref{sec:verification}. Comparison of the 
  reduced \ac{vda}-\ac{p} model against the \ac{hf} model.}}
  \label{fig:verification_p}
\end{figure}

\added{Further experiments were carried out with the \ac{vda}-\ac{pp} model to 
assess the other \ac{vda} discrete forms in Tab.~\ref{tab:vda_equations}. 
Taking the same parameters as with the Dirichlet case, extra quantities $h_{\rm 
s/p}$ and $h_{\rm p/p}$ were not estimated, they were set to 1,000 
[W/m\textsuperscript{2}C] and 10 [W/m\textsuperscript{2}C]. Results in 
Fig.~\ref{fig:verification_pp_comp} show that more accurate discretizations, 
lead to better approximations of the thermal response, as expected.} 

\subsection{Verification and validation against physical experiments}
\label{sec:results}

\subsubsection{Experimental campaign}
\label{sec:experiments}

An experimental campaign took place at the Monash Centre for Additive 
Manufacturing (MCAM) in Melbourne, Australia, with the purpose of (1) 
calibrating experimentally the thermal analysis FE framework described in 
Sect.~\ref{sec:formulation} and (2) assess the novel \ac{vda} model 
presented in Sect.~\ref{sec:virtual}.

The printing system employed for the experiments is the EOSINT M280 from 
Electro Optical Systems (EOS) GmbH. It uses an Yb-fibre laser with variable 
beam width and power up to 400 [W]. The printing process is carried out in a 
closed 250x250x325 $[\mathrm{mm}^3]$ chamber subject to a laminar flow of 
argon that prevents oxidation.

\begin{figure}[!h]
  \centering
  \begin{subfigure}[t]{0.48\textwidth}
    \begin{center}
	    \resizebox{0.9\textwidth}{!}{

\begin{tikzpicture}[scale=2]
    \filldraw[fill=teal!40!white] (0,0) rectangle (2.52,2.52);
    \fill[teal!80!white] (1.50,1.10) rectangle (1.80,1.40);
    \draw (1.10,1.10) rectangle (1.80,1.40);
    \draw[dashed] (1.50,1.10) -- (1.50,1.40);
    \draw[thin,<->] (0,-0.1) --node[below, font=\tiny]{252} (2.52,-0.1);
    \draw[thin,<->] (-0.1,0.0) --node[left, font=\tiny]{111} (-0.1,1.1);
    \draw[thin,<->] (-0.1,1.1) --node[left, font=\tiny]{30} (-0.1,1.4);
    \draw[thin,<->] (-0.1,1.4) --node[left, font=\tiny]{111} (-0.1,2.52);
    \draw[thin,<->] (1.9,1.1) --node[right, font=\tiny]{30} (1.9,1.4);
    \draw[thick,->] (0.0,0.0) -- (0.2,0.0) node[anchor=south, font=\tiny]{x};
    \draw[thick,->] (0.0,0.0) -- (0.0,0.2) node[anchor=west, font=\tiny]{y};
\end{tikzpicture}
	    }
	    \caption{Plane XY view.}
    \end{center}
  \end{subfigure}
  \quad
  \begin{subfigure}[t]{0.48\textwidth}
    \begin{center}
	    \resizebox{0.9\textwidth}{!}{

\begin{tikzpicture}[scale=2]
    \filldraw[fill=teal!40!white] (0,0) rectangle (2.52,0.45);
    \filldraw[fill=teal!40!white] (1.40,0.45) -- (1.80,0.85) -- (1.50,0.85) -- (1.10,0.45);
    \fill[teal!80!white] (1.6058,0.6558) -- (1.80,0.85) -- (1.50,0.85) -- (1.3058,0.6558) -- cycle;
    \draw[dashed] (1.3058,0.6558) -- (1.6058,0.6558);
    \draw (1.6058,0.6558) -- (1.80,0.85) -- (1.50,0.85) -- (1.3058,0.6558);
    \draw[thin,<->] (0.0,-0.1) --node[below, font=\tiny]{111} (1.1,-0.1);
    \draw[thin,<->] (1.1,-0.1) --node[below, font=\tiny]{30} (1.4,-0.1);
    \draw[thin,<->] (1.4,-0.1) --node[below, font=\tiny]{40} (1.8,-0.1);
    \draw[thin,<->] (1.8,-0.1) --node[below, font=\tiny]{71} (2.52,-0.1);
    \draw[thin,<->] (2.62,0.0) --node[right, font=\tiny]{45} (2.62,0.45);
    \draw[thin,<->] (2.62,0.45) --node[right, font=\tiny]{20.58} (2.62,0.6558);
    \draw[thin,<->] (2.62,0.6558) --node[right, font=\tiny]{19.42} (2.62,0.85);
    \draw[thin,<->] (1.50,0.95) --node[above, font=\tiny]{30} (1.80,0.95);
    \draw (1.5,0.45) arc[radius = 1mm, start angle= 0, end angle= 45] node[right, font=\tiny]{45$^\circ$};
    \draw[thick,->] (0.0,0.0) -- (0.2,0.0) node[anchor=south, font=\tiny]{x};
    \draw[thick,->] (0.0,0.0) -- (0.0,0.2) node[anchor=west, font=\tiny]{z};
\end{tikzpicture}
	    }
	    \caption{Plane XZ view.}
    \end{center}
  \end{subfigure} 
  \caption{\ac{mcam} experiment. Base plate and printed specimen 
  (mm). The simulated region is highlighted in dark teal.}
  \label{fig:geometry}
\end{figure}
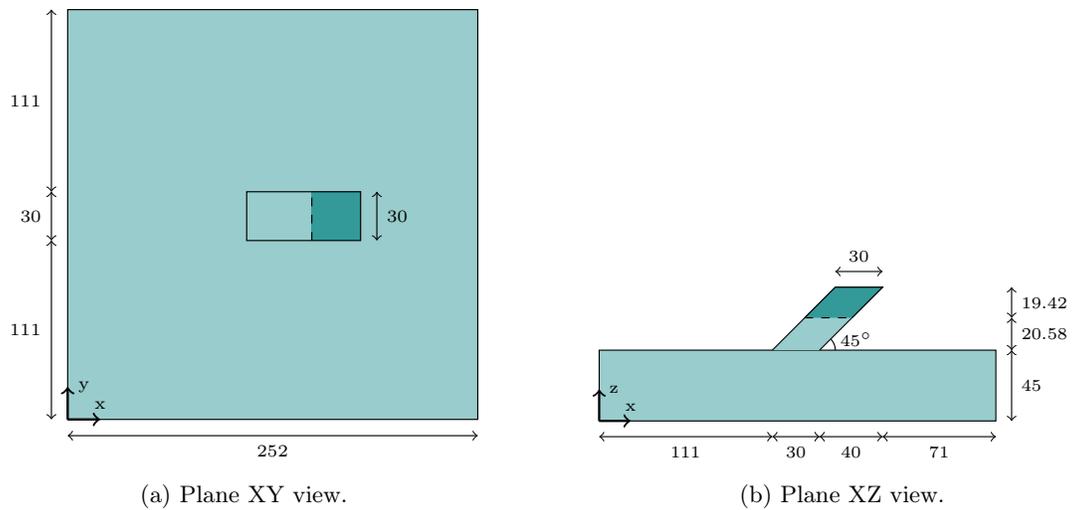

The printed specimen is an oblique square prism with the lower base located 
in the centre of the building plate and a 45-degrees slope, as shown in 
Fig.~\ref{fig:geometry}. Cross section dimensions are 30x30 $[\mathrm{mm}^2]$ 
and the height is 80 [mm].

\begin{figure}[!h]
  \centering
  \begin{subfigure}[t]{0.48\textwidth}
    \begin{center}
	    \resizebox{0.8\textwidth}{!}{

\begin{tikzpicture}
  \fill[teal!40!white] (6.6,6.6) rectangle (10.8,8.4);
  \fill[teal!80!white] (9.0,6.6) rectangle (10.8,8.4);
  \draw (6.6,6.6) rectangle (10.8,8.4);
  \draw[dashed] (9.0,6.6) -- (9.0,8.4);
  \filldraw (7.74,7.5) circle [radius=0.02] node[anchor=east,font=\tiny]{CH1,CH2};
  \filldraw (7.74,7.2) circle [radius=0.02] node[anchor=east,font=\tiny]{CH3};
  \filldraw (8.58,7.5) circle [radius=0.02] node[anchor=east,font=\tiny]{CH4};
  \filldraw (9.42,7.5) circle [radius=0.02] node[anchor=west,yshift=-0.02cm,font=\tiny]{CH5,CH6};
  \filldraw (9.39,7.5) circle [radius=0.02] node[anchor=east,font=\tiny]{CH7};
  \filldraw (9.39,7.2) circle [radius=0.02] node[anchor=east,font=\tiny]{CH8};
  \draw[thick,->] (6.6,6.6) -- (7.0,6.6) node[anchor=south, font=\tiny]{x};
  \draw[thick,->] (6.6,6.6) -- (6.6,7.0) node[anchor=west, font=\tiny]{y};
\end{tikzpicture}
	    }
	    \caption{Plane XY view.}
    \end{center}  
  \end{subfigure}
  \quad
  \begin{subfigure}[t]{0.48\textwidth}
    \begin{center}
	    \resizebox{0.8\textwidth}{!}{

\begin{tikzpicture}
  \filldraw[fill=teal!40!white] (8.40,2.7) -- (10.80,5.1) -- (9.00,5.1) -- (6.60,2.7) -- cycle;
  \fill[teal!80!white] (9.6348,3.9348) -- (10.80,5.1) -- (9.00,5.1) -- (7.8348,3.9348) -- cycle;
  \draw[dashed] (7.8348,3.9348) -- (9.6348,3.9348);
  \draw (9.6348,3.9348) -- (10.80,5.1) -- (9.00,5.1) -- (7.8348,3.9348);
  \filldraw (7.74,3.60) circle [radius=0.02] node[anchor=east,font=\tiny]{CH1};
  \filldraw (7.74,3.78) circle [radius=0.02] node[anchor=east,font=\tiny]{CH2,CH3};
  \filldraw (8.58,3.78) circle [radius=0.02] node[anchor=east,font=\tiny]{CH4};
  \filldraw (9.42,3.69) circle [radius=0.02] node[anchor=west,yshift=-0.05cm,font=\tiny]{CH5};
  \filldraw (9.42,3.60) circle [radius=0.02] node[anchor=west,yshift=-0.13cm,font=\tiny]{CH6};
  \filldraw (9.39,3.78) circle [radius=0.02] node[anchor=west,yshift=0.01cm,font=\tiny]{CH7,CH8};
  \draw[thick,->] (6.6,2.7) -- (7.0,2.7) node[anchor=south, font=\tiny]{x};
  \draw[thick,->] (6.6,2.7) -- (6.6,3.1) node[anchor=west, font=\tiny]{z};
\end{tikzpicture}
	    }
	    \caption{Plane XZ view.}
    \end{center}
  \end{subfigure} \\
  \vspace{0.2cm}
  \begin{subfigure}[t]{0.48\textwidth}
    \begin{center}
      \includegraphics[width=\textwidth]{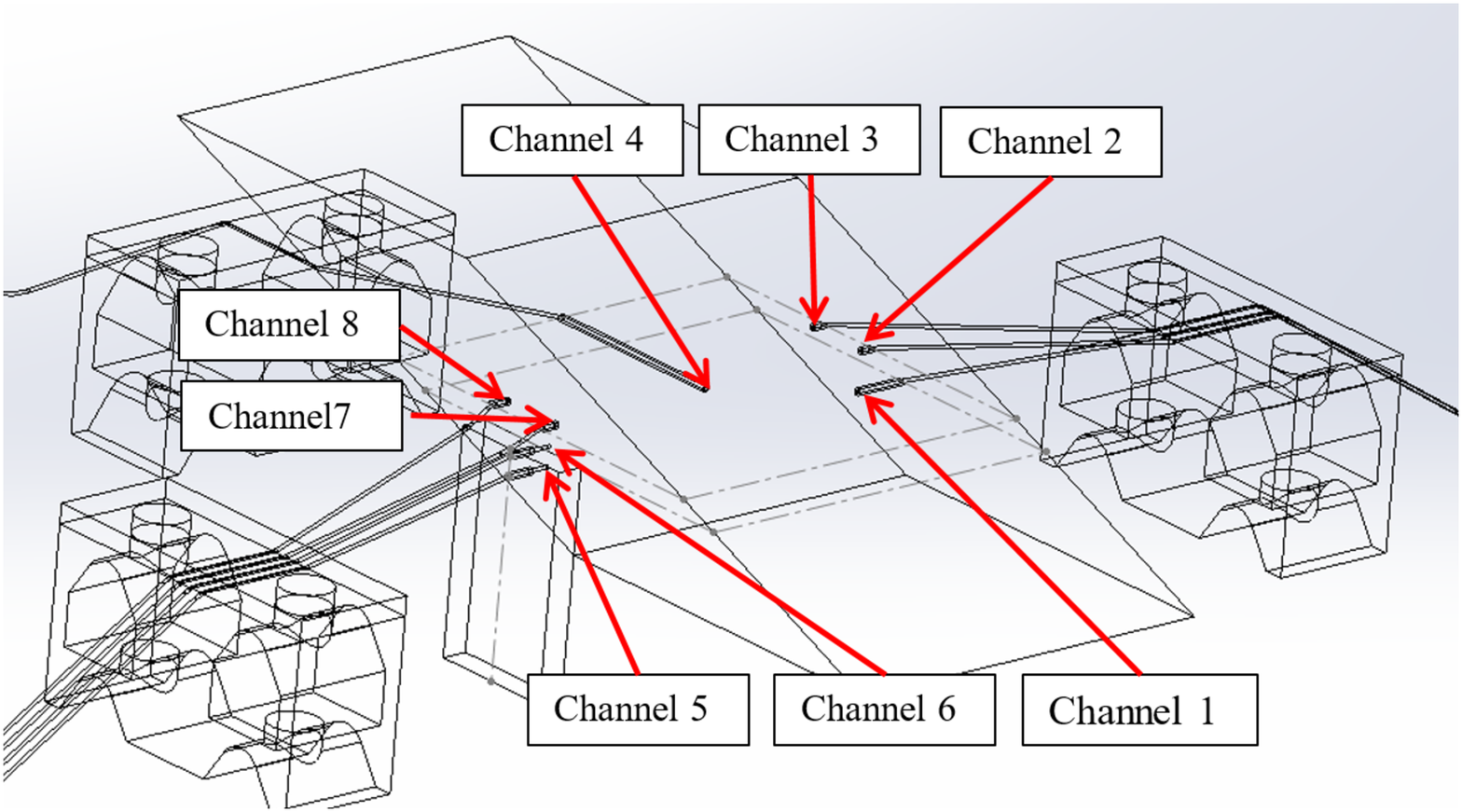}
      \caption{CAD view.}
    \end{center}
  \end{subfigure}
  \caption{Location of thermocouples at the specimen. The simulated region is 
  highlighted in dark teal.}
  \label{fig:thermocouples}
\end{figure}

\begin{figure}[!h]
  \centering
  \includegraphics[width=0.48\textwidth]{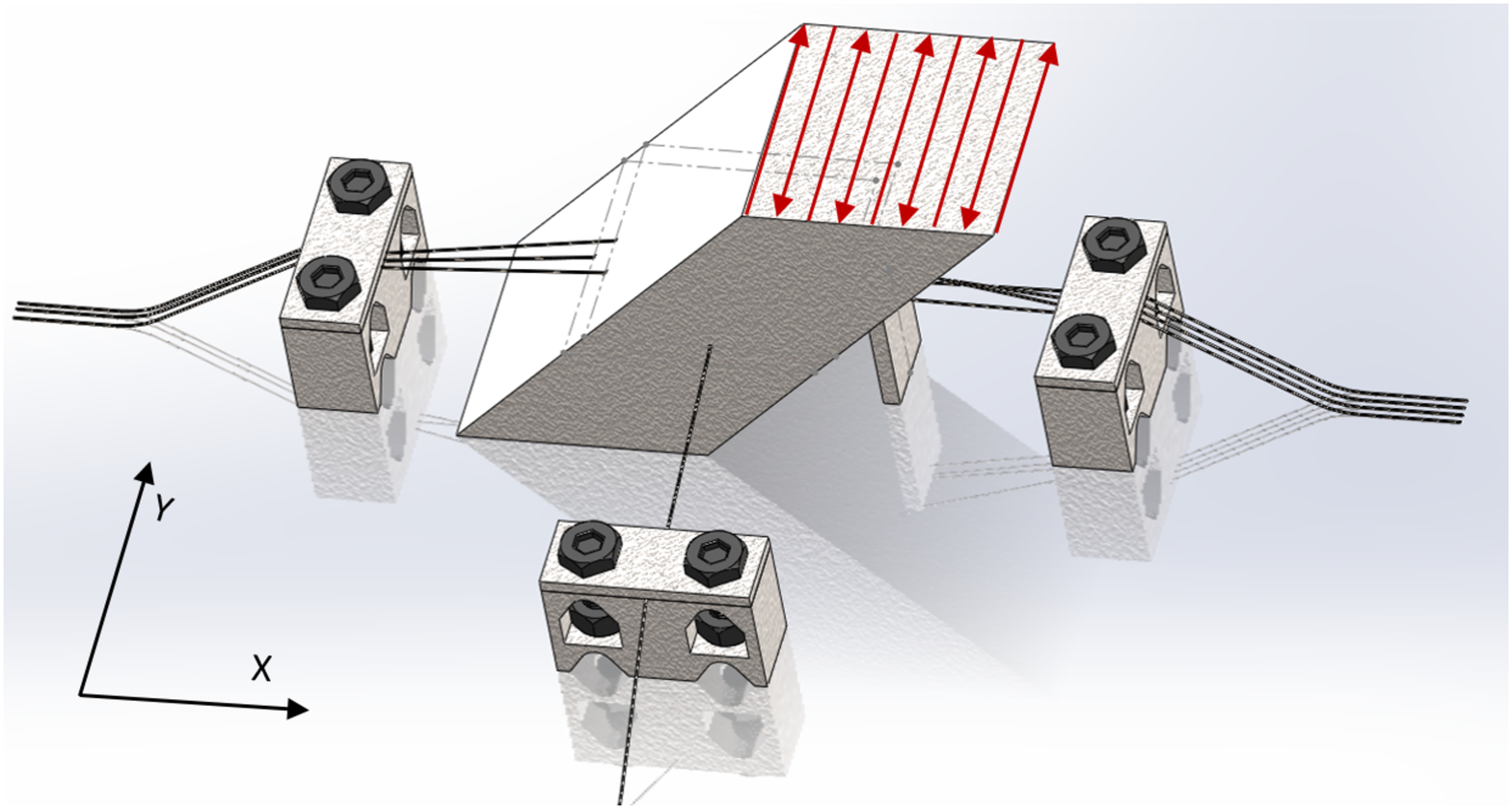}
  \caption{CAD view of thermocouple supports and orientative scanning path.}
  \label{fig:scanningpath}
\end{figure}

Eight thermocouples for temperature measurements are inserted into 0.78 [mm] 
diameter holes at the upward and downward facing lateral surfaces of the 
prism (CH1-4 and CH7-8) or in the powder bed (CH5-6). The position of the 
channels is indicated in Tab.~\ref{tab:coordinates} and 
Fig.~\ref{fig:thermocouples}. The printing job was interrupted at 20.58 [mm] 
height to install the thermocouples on a set of supporting structures that 
were printed together with the prism, as shown in 
Fig.~\ref{fig:scanningpath}.

K-type thermocouples and a Graphtec GL-900 8 high-speed data-logger are used 
for data gathering. The sampling rate of the data logger is 1 $[\mathrm{ms}]$ 
and the time constant of the thermocouples is 7 $[\mathrm{ms}]$. As 
thermocouples are not welded and can move inside the hole, their measurements 
can be perturbed.

\begin{table}[!h]
    \centering
    \begin{tabular}{ | c | c | }
        \hline
        Channel & (x,y,z)    \\
        \hline
        CH1 & (129,125,60)   \\
        \hline
        CH2 & (129,125,63)   \\
        \hline
        CH3 & (129,120,63)   \\
        \hline
        CH4 & (143,125,63)   \\
        \hline
        CH5 & (157,125,61.5) \\
        \hline
        CH6 & (157,125,60)   \\
        \hline
        CH7 & (156.5,125,63) \\
        \hline
        CH8 & (156.5,120,63) \\
        \hline
    \end{tabular}
    \vspace{0.2cm}
    \caption{Coordinates (mm) of the thermocouples with respect to the origin 
    of coordinates in Figure~\ref{fig:geometry}.}
    \label{tab:coordinates}
\end{table}

The process parameters used for the printing job are described in 
Tab.~\ref{tab:process}. The layer thickness is set to 30 $[\upmu 
\mathrm{m}]$. This means that 647 layers are deposited in about 3.5 [h] to 
build the samples. As observed, the recoat and laser relocation time varies 
between odd and even layers.

\begin{table}[!h]
    \centering
    \begin{tabular}{ | c | c c | }
        \hline
        Power input & $280$ & $[\mathrm{W}]$ \\
        \hline
        Scanning speed & $1,200$ & $[\mathrm{mm}/\mathrm{s}]$ \\
        \hline
        Layer thickness & $30$ & $[\upmu \mathrm{m}]$ \\
        \hline
        Hatch distance & $140$ & $[\upmu \mathrm{m}]$ \\
        \hline
        Beam offset & $15$ & $[\upmu \mathrm{m}]$ \\
        \hline
        Recoat time (odd layers) & $14.3$ & $[\mathrm{s}]$ \\
        \hline
        Recoat time (even layers) & $10.7$ & $[\mathrm{s}]$ \\
        \hline
    \end{tabular}
    \vspace{0.2cm}
    \caption{\ac{mcam} experiment. Process parameters adopted by the EOS 
    Machine.}
    \label{tab:process}
\end{table}

Regarding the scanning strategy, the laser travels along the y direction back 
and forth for each layer, as shown in Fig.~\ref{fig:scanningpath}. Note 
that the number of hatches drawn does not correspond to the actual number of 
hatches, which is a much higher value, according to the laser beam size.

The printed samples are made of Ti6Al4V Titanium alloy. The 
temperature-dependent properties of the bulk material, covering the range 
from room temperature to fusion temperature, are available 
in~\cite{chiumenti_neiva_2017}. The base plate is made of CP Ti, a material 
with similar thermal properties as those of Ti64. Complementary experiments 
in~\cite{chiumenti_neiva_2017} estimated the relative density of Ti64 powder 
at around 54 \% of the density of the bulk material at room temperature.

\begin{figure}[!h]
  \centering
  \resizebox{0.48\textwidth}{!}{
  \input{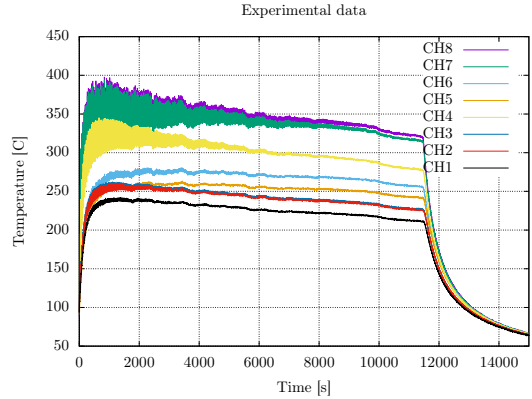}
  }
  \caption{Temperature data gathered at the eight thermocouple channels. The 
  evolution is initially ascendant and very oscillatory, but then it 
  stabilizes and decreases slowly. After the printing, temperature drops to 
  room temperature.}
  \label{fig:data}
\end{figure}

Fig.~\ref{fig:data} describes the experimental data gathered at the eight 
thermocouple channels. During the printing of the first layers, the 
thermocouples are close to the laser and a sharp and highly oscillatory 
temperature build-up takes place. This trend stabilizes a little before the 
hundredth layer, when the thermocouples are far enough from the laser spot. 
Then, the temperature evolution enters a slowly-cooling quasi steady-state 
regime, until the process finishes and the temperature falls to room 
temperature.

\subsubsection{Methodology}
\label{subsec:methodology}

The numerical experiments were supported by a framework consisting of: (1) 
\href{http://www.fempar.org/}{FEMPAR}~\cite{badia2017fempar}, an advanced 
object-oriented parallel FE library for large scale computational science and 
engineering, (2) \href{https://dakota.sandia.gov/}{Dakota}~\cite{dakota}, a 
suite of iterative mathematical and statistical methods for parameter 
estimation, sensitivity analysis and optimization of computational models, 
and (3) 
\href{https://caminstech.upc.edu/es/calculintensiu}{TITANI}~\ref{tab:titani}, 
a High Performance Computer (HPC) cluster at the Universitat Politècnica de 
Catalunya (Barcelona, Spain).

Such advanced computational framework, integrating FEA tools with scientific 
software for parameter exploration on a HPC platform, has not been previously 
observed in the literature related to the numerical simulation of metal AM 
processes, but it is very convenient to carry out the verification and 
validation tasks at the scale of the experiment.

\begin{table}[!h]
    \centering
    \begin{tabular}{ | c | c | }
        \hline
        Nodes      & 5 DELL PowerEdge R630              \\ \hline
        CPUs       & 2 x Intel Xeon E52650L v3 (1.8GHz) \\ \hline
        Cores      & 12 cores per processor             \\ \hline
        Cache      & 30 MB                              \\ \hline
        RAM        & 256 GB                             \\ \hline
        Local disk & 2 x 250 GB SATA                    \\ \hline
    \end{tabular}
    \vspace{0.2cm}
    \caption{Overview of the architecture of TITANI.}
    \label{tab:titani}
\end{table}

An overview of the procedure followed during the numerical tests is:
\begin{enumerate}
	\item Implementation of the computational model described in  
	Sect.~\ref{sec:formulation} and the \ac{vda} model in 
	Sect.~\ref{sec:virtual} in FEMPAR.
	\item Implementation of an interface to communicate Dakota with FEMPAR.
	\item Design and implementation of a physically accurate reference 
	(\ac{hf}) heat transfer model.
	\item Calibration and validation of the \ac{hf} model against 
	experimental data generated at the MCAM research centre.
	\item Repeat (3) and (4) for two additional \ac{htc}-\ac{pp} and 
	\ac{vda}-\ac{pp} reduced HPC models in TITANI. Comparison of the 
	reduced-domain variants with the full-domain \ac{hf} model.
\end{enumerate}

According to this, three different numerical models were tested, the only 
difference among them being how heat loss through the powder bed is accounted 
for. In the \ac{hf} model, the purpose is to maximize the accuracy of the 
model, by including the powder bed into the computational domain of analysis, 
to establish a numerical reference for the reduced models.

In the \ac{htc}-\ac{pp} model, the powder-bed is excluded from the 
computational domain and heat loss through the powder bed is modelled with a 
constant heat conduction boundary condition at the solid-powder contact 
surfaces. The \ac{vda}-\ac{pp} model adopts the same hypotheses of the 
\ac{htc}-\ac{pp} model, except for the computation of the \ac{htc} at the 
solid-powder interfaces, which is derived from Eq.~\eqref{eq:vda_equivalent}, 
considering a single quadratic element \added{(cf. (1E-Q2) in 
Tab.~\ref{tab:vda_equations})} to solve the \ac{vda} 1D heat conduction problem.

The calibration is done w.r.t. the measurements in thermocouple channels CH2 
and CH4, separated 14 mm horizontally. {Only measurements during printing 
are considered; the cooling stage is simulated, but not calibrated.} The 
numerical response is then compared with the experimental data of CH8, one of 
the furthest from CH2 and CH4, as a way to validate the model. Remaining 
thermocouples are either close to CH2, CH4, CH8, or outside the sample.

An important feature is that both \ac{vda}-\ac{pp} and \ac{htc}-\ac{pp} 
models inherit all the simulation data (process parameters, material 
properties,...) from the \ac{hf} model and only the relevant parameters of 
each model are estimated. Tab.~\ref{tab:comparison} gathers all the 
simulation data from the three models and highlights the calibration 
parameters of the reduced models.

\subsubsection{Calibration of the powder-bed model}
\label{subsec:calibration}

The reference model that will be used later to assess the accuracy and 
performance of the \ac{vda} model must reproduce as closely as possible the 
physical process of metal deposition. Likewise, the size of the simulation 
must be chosen to enable a full sensitivity analysis and iterative parameter 
estimation in reasonable computational times.

A locally accurate simulation of the metal deposition process must include 
the powder-bed in the computational domain, the FE mesh must conform to the 
printed layers, the mesh size must be smaller than the laser beam spot size, 
and the scanning path must be tracked element by element.

However, complying with these requirements is not always possible from the 
computational point of view. For instance, in this experiment, assuming a 
uniform mesh with element size 50x50x30 $[\upmu \mathrm{m}^3]$, a single 
layer of the specimen would be composed of 360,000 elements to be printed in 
360,000 time steps.

Besides, the focus of this work is on the thermal analysis \textit{at the 
part scale}, as commented in Sect.~\ref{sec:introduction}. This allows us to 
control the problem size and the number of time steps, by relaxing the 
previous discretization requirements with suitable approximations.

The most relevant simplification in this work is the adoption of a 
layer-by-layer activation strategy, as in Sect.~\ref{sec:verification}. Even 
though this assumption is computationally very appealing, it sacrifices the 
local thermal history of the response. {However}, it 
allows us to recover average thermal responses, as discussed 
in~\cite{chiumenti_neiva_2017}, and enables parametric exploration of the 
computational model in moderate times. {In any case, regardless of the 
deposition sequence, the reference \ac{hf} model is, by construction, more 
accurate than the reduced-domain \ac{vda}-\ac{pp} and \ac{htc}-\ac{pp}  
versions.}

\begin{figure}[!h]
  \centering
  \begin{subfigure}[t]{0.45\textwidth}
    \centering
    \includegraphics[width=0.9\textwidth]{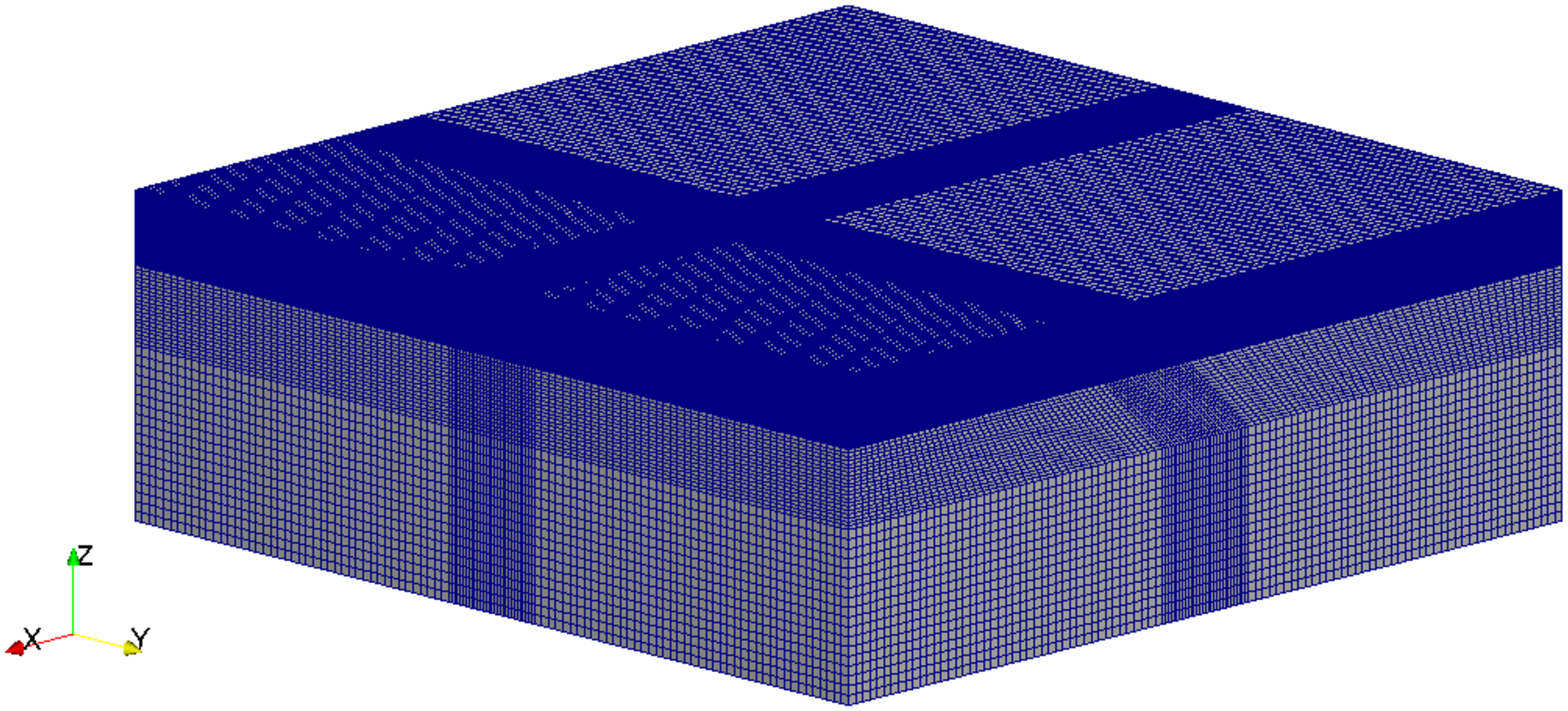}
    \caption{\ac{hf} model.}
    \label{fig:PBmesh}
  \end{subfigure}
  \quad
  \begin{subfigure}[t]{0.45\textwidth}
    \centering
    \includegraphics[width=0.9\textwidth]{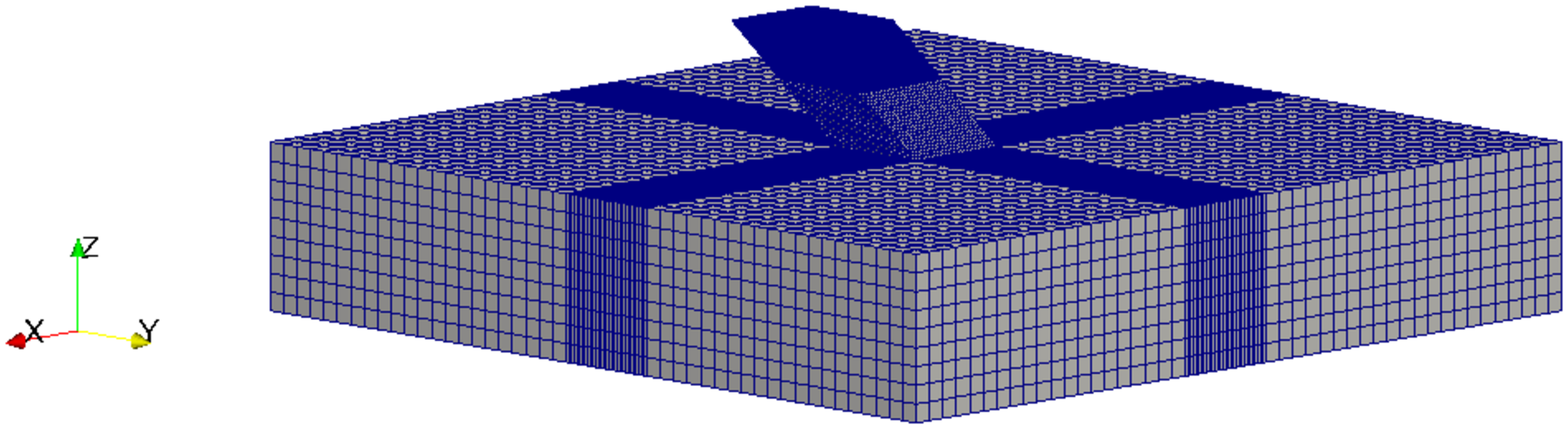}
    \caption{\ac{vda}-\ac{pp} and \ac{htc}-\ac{pp} models.}
    \label{fig:VLPmesh}
  \end{subfigure}
  \caption{FE meshes used in the analysis of the \ac{mcam} experiment. They 
  conform to the printed layers, that is the reason for the element 
  concentration in the z-direction.}
  \label{fig:FEmeshes}
\end{figure}

As a result of the previous considerations, the FE discretization is a 
structured mesh of 9,565,788 hexahedral elements and 9,742,768 nodes. As 
observed in Fig.~\ref{fig:PBmesh}, the FE mesh is a box containing the 
printed specimen, the building plate, and the powder bed. Supporting 
structures of the thermocouples are not included in the analysis, because 
they are far and small enough to have little influence in the results at the 
specimen.

\added{A static mesh is employed throughout the simulation.} \replaced{Element 
size was determined with a convergence test and it varies, depending on the 
region of the model.}{Mesh size varies, depending on the region of the model, 
and was fixed after a mesh convergence test.} Fine 1x1x0.03 
[mm\textsuperscript{3}] elements are prescribed at the 
\replaced{simulated}{printed} region, while larger mesh sizes are specified 
\replaced{elsewhere}{at the remaining regions}. Note that, as a result of the 
layer-averaging in time and using a uniform heat source distribution, the mesh 
size no longer needs to be smaller than the laser spot size.

The numerical simulation begins after placing the thermocouples and resuming 
the job. It continues with the deposition of the remaining 647 layers and 
finishes with the cooling of the whole ensemble. This amounts to almost 
13,000 [s] of deposition process.

All simulation data is listed in Tab.~\ref{tab:comparison}. Some comments 
arise on the parameter values: 
\begin{enumerate}
	\item The heat absorption coefficient, estimated at 70 \%, is the most 
	sensitive parameter of the model, as also observed 
	in~\cite{chiumenti_numerical_2017}. 
	\item The temperature-dependent values of the powder thermal conductivity 
	in Tab.~\ref{tab:powder_conductivity} were estimated using 
	Eq.~\ref{eq:powderconductivity}.
	\item Temperatures of air, powder and plate were measured in the 
	chamber. They suffer variations throughout the process, but the numerical 
	model is barely sensitive to them, with the exception of plate 
	temperature. For this quantity, an estimated evolution law is assumed in
	Tab.~\ref{tab:platform}, taking into account that the building plate is
	heated during the printing phase.
	\item The dimensions of the numerical experiment (mesh size and number of 
	time steps) can only be appropriately dealt with parallel computing 
	techniques. Still, the calibration procedure was rather slow. For 
	instance, using 20 CPUs of TITANI, the execution time of a single 
	evaluation of the \ac{hf} model was approximately 50 [h].
\end{enumerate}

\begin{table}[!h]
    \centering
    \begin{tabular}{ | c | c | c | c | c | }
        \hline
        \textbf{Parameter} & \textbf{\ac{hf} model} & 
        \textbf{\ac{htc}-\ac{pp} model} & \textbf{\ac{vda}-\ac{pp} model} & 
        \textbf{Units} \\ \hline
        \multicolumn{5}{|c|}{\textbf{Process parameters}} \\ \hline
        Layer thickness & \multicolumn{3}{ c |}{30} & $\upmu$m \\ \hline
        Scanning speed & \multicolumn{3}{ c |}{5.6} & mm/s \\ \hline
        Backward speed (odd layer) & \multicolumn{3}{ c |}{2.8} & mm/s \\ 
        \hline
        Backward speed (even layer) & \multicolumn{3}{ c |}{2.1} & mm/s \\ 
        \hline
        Laser power & \multicolumn{3}{ c |}{280} & W \\ \hline
        Laser efficiency & \multicolumn{3}{ c |}{70} & \% \\ \hline
        \multicolumn{5}{|c|}{\textbf{Material properties}} \\ \hline
        Bulk thermal properties & 
        \multicolumn{3}{ c |}{In~\cite{chiumenti_neiva_2017}} & - \\ \hline
        Powder thermal properties & In Tab.~\ref{tab:powder_conductivity} & - 
        & In Tab.~\ref{tab:powder_conductivity} & - \\ \hline
        \multicolumn{5}{|c|}{\textbf{Boundary conditions}} \\ \hline
        HTC air & \multicolumn{3}{ c |}{10} & W/m\textsuperscript{2}C        
        \\ \hline
        Air temperature & \multicolumn{3}{ c |}{35} & C \\ \hline
        HTC at chamber wall & \multicolumn{3}{ c |}{10} & 
        W/m\textsuperscript{2}C \\ \hline
        Temperature at chamber wall & \multicolumn{3}{ c |}{93} & 
        C \\ \hline
        HTC at plate & \multicolumn{3}{ c |}{10} & 
        W/m\textsuperscript{2}C \\ \hline
        Platform temperature  & \multicolumn{3}{ c |}{In 
        Tab.~\ref{tab:platform}} & - \\ \hline
        Equivalent HTC solid-powder & - & {\color{blue}{21}} & - & 
        W/m\textsuperscript{2}C \\ \hline
        Powder temperature & \multicolumn{3}{ c |}{93} & C \\ \hline
        Initial temperature & \multicolumn{3}{ c |}{93} & C \\ \hline
        \multicolumn{5}{|c|}{\textbf{Virtual powder model 
        data}} \\ \hline
        Thickness & - & - & {\color{blue}{36}} & mm \\ \hline
        HTC solid-VDA & - & - & {\color{blue}{4,150}} & 
        W/m\textsuperscript{2}C \\ \hline
        HTC VDA-powder & - & - & 1,000 & W/m\textsuperscript{2}C \\ \hline
        Volume factor & - & - & 1 & - \\ \hline
        Surface factor & - & - & 1 & - \\ \hline
        \multicolumn{5}{|c|}{\textbf{Problem size}} \\ \hline
        Mesh size & 9,742,768 & \multicolumn{2}{ c |}{696,199} & nodes \\ 
        \hline
        Number of layers & \multicolumn{3}{ c |}{647} & layers \\ \hline
        Number of time steps & \multicolumn{3}{ c |}{1,434} & steps \\ \hline
        Activation strategy & \multicolumn{3}{ c |}{layer-by-layer} & - \\ 
        \hline
        \multicolumn{5}{|c|}{\textbf{Computational cost}} \\ \hline
        {Number of CPUs} & \multicolumn{3}{ c |}{{20}} & 
        {CPUs} \\ \hline
        Execution time & 50 & \multicolumn{2}{ c|}{{3}} & h \\ 
        \hline
    \end{tabular}
    \vspace{0.2cm}
    \caption{Comparison of thermal models analysed in the \ac{mcam} 
    experiment. Calibration parameters of the \ac{htc}-\ac{pp} and 
    \ac{vda}-\ac{pp} models marked in blue. Boundary conditions are also 
    described in Fig.~\ref{fig:boundary}.}
    \label{tab:comparison}
\end{table}

\begin{table}[!h]
    \centering
    \begin{tabular}{ | c | c | }
        \hline
        \textbf{Time [s]}  & \textbf{Temperature [C]} \\ \hline
                       0.0 &                     93.0 \\ \hline
                  11,470.0 &                     93.0 \\ \hline
                  12,000.0 &                     35.0 \\ \hline
                  15,000.0 &                     35.0 \\ \hline
    \end{tabular}
    \vspace{0.2cm}
    \caption{\ac{mcam} experiment. Estimated evolution of temperature at the 
    lower surface of the building plate.}
    \label{tab:platform}
\end{table}

\begin{table}[!h]
    \centering
    \begin{tabular}{ | c | c | }
        \hline
        \textbf{Temperature [C]}  & \textbf{Conductivity [W/mC]} \\ \hline
                             20.0 &                        0.288 \\ \hline
                            200.0 &                        0.407 \\ \hline
                            300.0 &                        0.466 \\ \hline
                            400.0 &                        0.520 \\ \hline
                            500.0 &                        0.573 \\ \hline
                            600.0 &                        0.630 \\ \hline
                            700.0 &                        0.684 \\ \hline
                            800.0 &                        0.746 \\ \hline
                            900.0 &                        0.808 \\ \hline
                          1,100.0 &                        0.904 \\ \hline
                          1,227.0 &                        0.976 \\ \hline
                          1,500.0 &                        1.115 \\ \hline
                          1,600.0 &                        1.168 \\ \hline
                          1,660.0 &                        1.252 \\ \hline
    \end{tabular}
    \vspace{0.2cm}
    \caption{Temperature-dependent thermal conductivity of the Ti64 powder, 
    according to Eq.~\eqref{eq:powderconductivity}.}
    \label{tab:powder_conductivity}
\end{table}

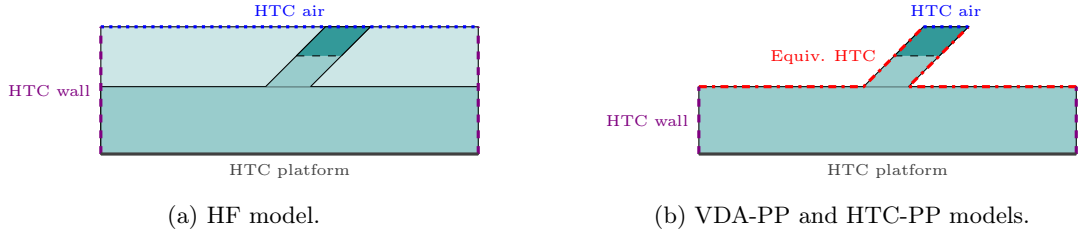
\begin{figure}[!h]
  \centering
  \begin{subfigure}[t]{0.48\textwidth}
    \begin{center}
	    \resizebox{0.9\textwidth}{!}{

\begin{tikzpicture}[scale=2]
    \filldraw[fill=teal!20!white] (0,0) rectangle (2.52,0.85);
    \filldraw[fill=teal!40!white] (0,0) rectangle (2.52,0.45);
    \filldraw[fill=teal!40!white] (1.40,0.45) -- (1.80,0.85) -- (1.50,0.85) -- (1.10,0.45);
    \fill[teal!80!white] (1.6058,0.6558) -- (1.80,0.85) -- (1.50,0.85) -- (1.3058,0.6558) -- cycle;
    \draw[dashed] (1.3058,0.6558) -- (1.6058,0.6558);
    \draw (1.6058,0.6558) -- (1.80,0.85) -- (1.50,0.85) -- (1.3058,0.6558);
    \draw[very thick,dotted,color=blue] (0.00,0.85) -- node[above, font=\tiny]{HTC air}      (2.52,0.85);
    \draw[very thick,dashed,color=violet] (0.00,0.00) -- node[left, font=\tiny]{HTC wall} (0.00,0.85);
    \draw[very thick,dashed,color=violet] (2.52,0.00) -- (2.52,0.85);
    \draw[very thick,color=black!50!gray] (0.00,0.00) -- node[below, font=\tiny]{HTC platform} (2.52,0.00);
\end{tikzpicture}
	    }
	    \caption{\ac{hf} model.}
    \end{center}
  \end{subfigure}
  \quad
  \begin{subfigure}[t]{0.48\textwidth}
    \begin{center}
	    \resizebox{0.9\textwidth}{!}{

\begin{tikzpicture}[scale=2]
    \filldraw[fill=teal!40!white] (0,0) rectangle (2.52,0.45);
    \filldraw[fill=teal!40!white] (1.40,0.45) -- (1.80,0.85) -- (1.50,0.85) -- (1.10,0.45);
    \fill[teal!80!white] (1.6058,0.6558) -- (1.80,0.85) -- (1.50,0.85) -- (1.3058,0.6558) -- cycle;
    \draw[dashed] (1.3058,0.6558) -- (1.6058,0.6558);
    \draw (1.6058,0.6558) -- (1.80,0.85) -- (1.50,0.85) -- (1.3058,0.6558);
    \draw[very thick,dashdotted,color=red] (0.00,0.45) -- (1.10,0.45) -- node[left, font=\tiny, xshift=-0.1cm]{Equiv. HTC} (1.50,0.85);
    \draw[very thick,dashdotted,color=red] (2.52,0.45) -- (1.40,0.45) -- (1.80,0.85);
    \draw[very thick,dotted,color=blue] (1.50,0.85) -- node[above, font=\tiny]{HTC air} (1.80,0.85);
    \draw[very thick,dashed,color=violet] (0.00,0.00) -- node[left, font=\tiny]{HTC wall} (0.00,0.45);
    \draw[very thick,dashed,color=violet] (2.52,0.00) -- (2.52,0.45);
    \draw[very thick,color=black!50!gray] (0.00,0.00) -- node[below, font=\tiny]{HTC platform} (2.52,0.00);
\end{tikzpicture}
	    }
	    \caption{\ac{vda}-\ac{pp} and \ac{htc}-\ac{pp} models.}
    \end{center}
  \end{subfigure} 
  \caption{Boundary conditions of \ac{mcam} experiment.}
  \label{fig:boundary}
\end{figure}

\begin{figure}[!h]
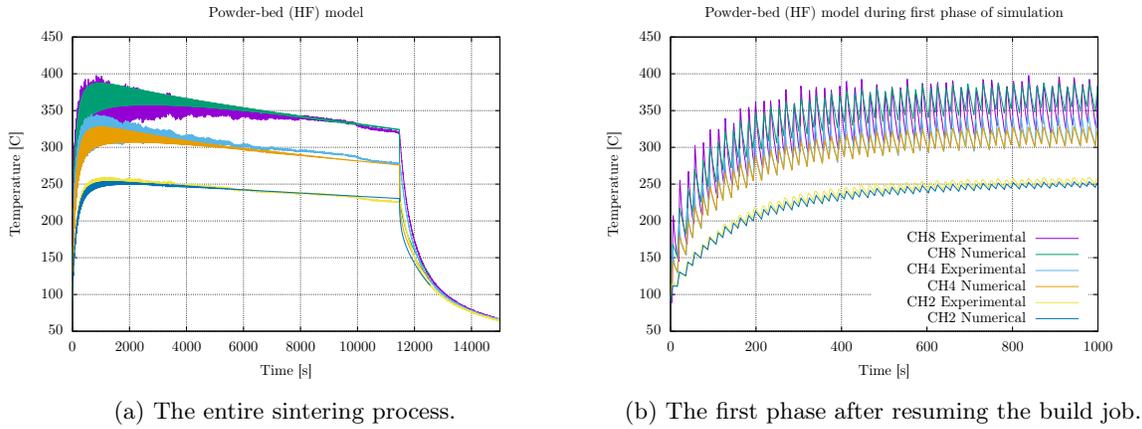

  \centering
  \begin{subfigure}[t]{0.48\textwidth}
    \scalebox{.55}{\input{MCAM-II-PB.tex}}
    \caption{The entire sintering process.}
    \label{fig:PBentire}
  \end{subfigure}
  \quad
  \begin{subfigure}[t]{0.48\textwidth}
    \scalebox{.55}{\input{MCAM-II-PB_closeUp.tex}}
    \caption{The first phase after resuming the build job.}
    \label{fig:PBcloseup}
  \end{subfigure}
  \caption{\ac{mcam} experiment. Numerical results of the \ac{hf} model. 
  Close agreement is observed for both calibration channels (CH2 and 
  CH4) and the validation channel (CH8).}
  \label{fig:PBresults}
\end{figure}

Fig.~\ref{fig:PBentire} {and Tab.~\ref{tab:error_experiments}} 
compare the numerical response with the experimental measurements. A 
very good agreement of predicted values and average rates in time can be 
observed during the whole printing process, both for the reference calibration 
channels, namely CH2 and CH4, and the validation channel CH8.

\begin{table}[!h]
    \centering
    \begin{tabular}{ | c | c | c | }
        \hline
        \textbf{Channel} & \textbf{MAE [C]} & \textbf{MRE [\%]} \\ \hline
        CH2 & 2.4 & 1.0 \\ \hline
        CH4 & 6.3 & 2.1 \\ \hline
        CH8 & 5.0 & 1.5 \\ \hline
    \end{tabular}
    \vspace{0.2cm}
    \caption{{\ac{mcam} experiment. Mean Absolute Error (MAE) and Mean 
    Relative Error (MRE) of the temperature evolution during the printing phase 
    between the experimental data and the calibrated \ac{hf} model.}}
    \label{tab:error_experiments}
\end{table}

\subsubsection{Assessment of the HTC-PP model}
\label{subsec:HTCmodel}

Exclusion of the powder-bed from the computational domain leads to a 
significant reduction in the size of the problem. In this case, the FE mesh 
(Fig.~\ref{fig:VLPmesh}) consists of 647,856 elements and 696,199 nodes. As 
a result, the \ac{htc}-\ac{pp} model can be solved with significantly less 
computational resources and time than the previous model, i.e. in 3 [h] using 
20 CPUs, and the parametric exploration is also much faster.

The HTC model is characterized by modelling heat transfer through the 
powder-bed with a constant heat conduction boundary condition on the 
solid-powder contact surfaces. For calibration against the experimental 
measurements, all the simulation data is the same as the one of the \ac{hf} 
model and the only variable is the HTC of the aforementioned boundary condition.

To interact with Dakota, the parameter estimation problem is formulated as an 
optimization problem of minimizing the least-squares error between data and 
variable-dependent response. The optimization problem is solved with a 
derivative-free local method (pattern search) until convergence.

To start the least-squares solver, the HTC can be initially estimated as 
\begin{equation}
    HTC(\T) = \frac{k_{\rm pwd}(\T)}{s_{\rm pwd}},
\end{equation}
\noindent{evaluated in the range of observed temperatures (200-400 [C]). 
Here, $k_{\rm pwd}$ is the conductivity of the powder and $s_{\rm pwd}$ is 
the virtual loose powder thickness of the \ac{vda}. Using this rule of thumb, 
the HTC is initially set at 13-17 [W/m\textsuperscript{2}C] and converges to 
21 [W/m\textsuperscript{2}C].}

\begin{figure}[!h]
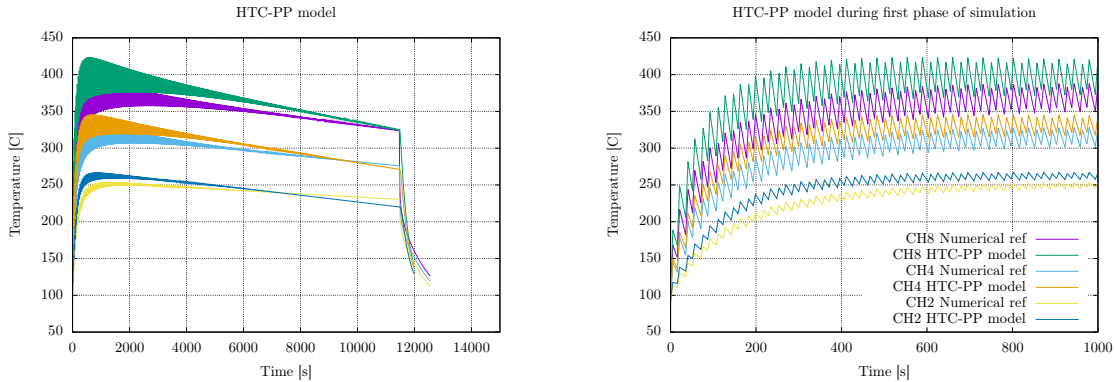

  \centering
  \begin{subfigure}[t]{0.48\textwidth}
    \scalebox{.55}{\input{MCAM-II-HTC.tex}}
    \caption{The entire sintering process.}
    \label{fig:HTCentire}
  \end{subfigure}
  \quad
  \begin{subfigure}[t]{0.48\textwidth}
    \scalebox{.55}{\input{MCAM-II-HTC_closeUp.tex}}
    \caption{The first phase after resuming the build job.}
    \label{fig:HTCcloseup}
  \end{subfigure}
  \caption{\ac{mcam} experiment. Numerical results of the \ac{htc}-\ac{pp} 
  model {against the reference \ac{hf} one}. In spite of the physical 
  simplifications, the maximum numerical error is still bounded by 15 \% at all 
  channels, but average temperature slopes in time are slightly different.}
  \label{fig:HTCresults}
\end{figure}

Concerning the numerical results, as seen in Fig.~\ref{fig:HTCresults} 
{and Tab.~\ref{tab:error_htc}}, the simplification of the physics 
(exclusion of powder-bed + equivalent boundary condition) has obvious negative 
consequences in the accuracy. {Besides, average rate of initial 
temperature build-up and cooling rate at the quasi-steady state regime are 
slightly different.} In spite of this, the {maximum} error of the 
\ac{htc}-\ac{pp} model with respect to the \ac{hf} {one} is 
bounded by 15 \%, even for the CH8 channel. 

\begin{table}[!h]
    \centering
    \begin{tabular}{ | c | c | c | c | c | }
        \cline{2-5}
        \multicolumn{1}{c|}{} & \multicolumn{2}{c|}{\textbf{\ac{hf} model}} 
        & \multicolumn{2}{c|}{\textbf{experiments}} \\ \hline
        \textbf{Channel} & \textbf{MAE [C]} & \textbf{MRE [\%]} & \textbf{MAE 
        [C]} & \textbf{MRE [\%]} \\ \hline
        CH2 & 6.7  & 2.8 & 4.5  & 1.9 \\ \hline
        CH4 & 8.3  & 2.8 & 5.5  & 1.8 \\ \hline
        CH8 & 15.0 & 4.3 & 18.8 & 5.4 \\ \hline
    \end{tabular}
    \vspace{0.2cm}
    \caption{{\ac{mcam} experiment. Mean Absolute Error (MAE) and Mean 
    Relative Error (MRE) of the temperature evolution during the printing phase 
    between the \ac{htc}-\ac{pp} model and the reference \ac{hf} or the 
    measured data.}}
    \label{tab:error_htc}
\end{table}

\subsubsection{Assessment of the VDA-PP model}
\label{subsec:VLPmodel}

\begin{figure}[!h]
  \centering
  \begin{subfigure}[t]{0.45\textwidth}
    \centering
    \includegraphics[width=0.9\textwidth]{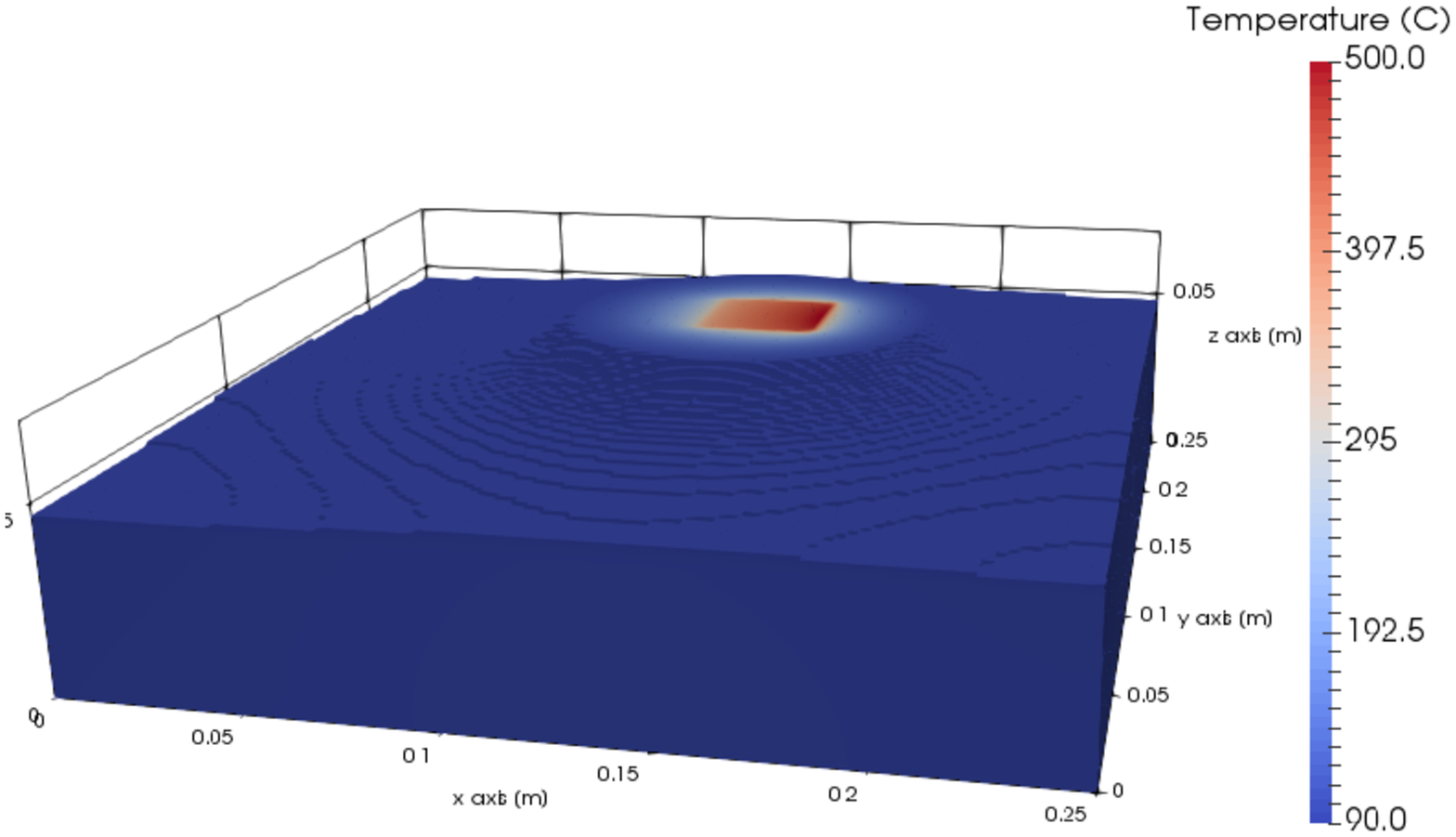}
    \caption{\ac{hf} model after filtering nodal values below the initial 
    temperature of the powder. This allows one to approximate the thickness 
    of the virtual powder.}
    \label{fig:PBcontour}
  \end{subfigure}
  \quad
  \begin{subfigure}[t]{0.45\textwidth}
    \centering
    \includegraphics[width=0.9\textwidth]{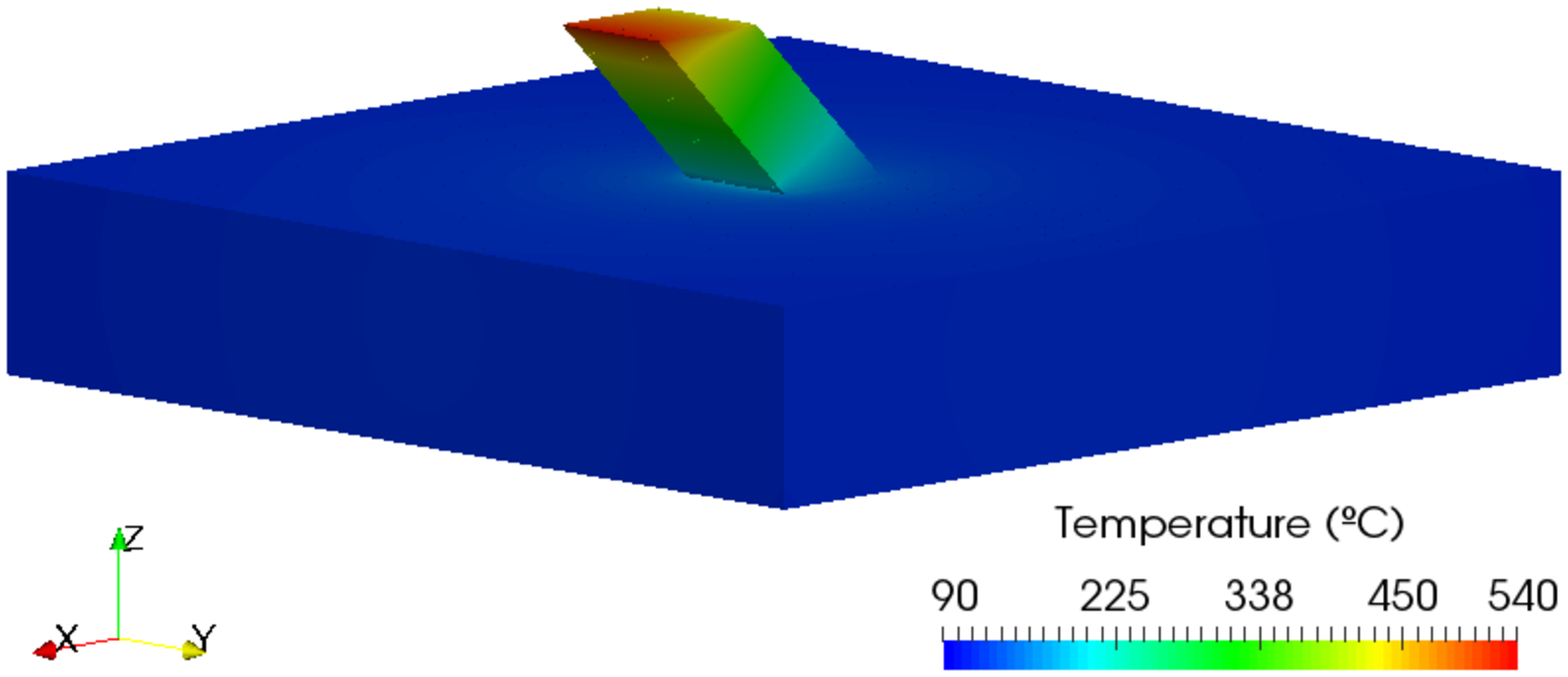}
    \caption{\ac{vda}-\ac{pp} model.}
    \label{fig:VLPcontour}
  \end{subfigure}
  \caption{\ac{mcam} experiment. Contour plots of temperatures of the \ac{hf} 
  and \ac{vda}-\ac{pp} models, represented at different time steps.}
  \label{fig:ContourPlots}
\end{figure}

As seen in the contour plot of temperatures in Fig.~\ref{fig:PBcontour}, if 
the nodal values below the initial temperature of the powder are filtered, it 
is exposed that thermal gradients concentrate around the printed region. 

Taking the FE mesh and the simulation data from the \ac{htc}-\ac{pp} model, 
the Dakota least-squares solver was reformulated for the \ac{vda}-\ac{pp} 
model. Some parameters of the \ac{vda}-\ac{pp} model can be fixed to reduce 
the number of calibration variables. For instance, the volume and surface 
factors $F_{\rm volume}$ and $F_{\rm surface}$ can be set to 1, according to 
the shell-like shape of the \ac{vda} region.

As a result of this, the experimental calibration variables are now the 
thickness of the \ac{vda} $s_{\rm pwd}$ and the HTC at the solid-\ac{vda} 
interface $h_{\rm s/p}$. The HTC at the \ac{vda}-powder interface $h_{\rm p/p}$ 
is also ruled out, after detecting low sensitivity to this parameter. However, 
its value should be large enough to have temperatures close to the initial 
temperature of the powder at the virtual boundary.

As for the initial approximations, from Figure~\ref{fig:PBcontour} the 
\ac{vda} thickness is set to 30 [mm], whereas the HTC solid-\ac{vda} is set 
to 3,000-4,000 [W/m\textsuperscript{2}C], which is in the range of typical 
values of HTC for metal-sand contact surfaces in sand casting.

\begin{figure}[!h]
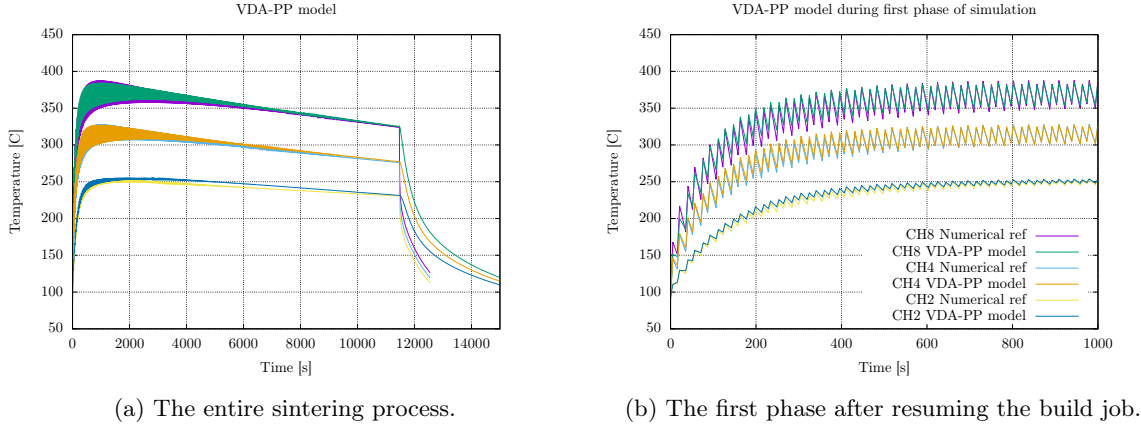

  \centering
  \begin{subfigure}[t]{0.48\textwidth}
    \scalebox{.55}{\input{MCAM-II-VM.tex}}
    \caption{The entire sintering process.}
    \label{fig:VLPentire}
  \end{subfigure}
  \quad
  \begin{subfigure}[t]{0.48\textwidth}
    \scalebox{.55}{\input{MCAM-II-VM_closeUp.tex}}
    \caption{The first phase after resuming the build job.}
    \label{fig:VLPcloseup}
  \end{subfigure}
  \caption{\ac{mcam} experiment. Numerical results of the \ac{vda}-\ac{pp} 
  model {against the reference \ac{hf} one}. \ac{vda}-\ac{pp} clearly 
  improves the \ac{htc}-\ac{pp}, in the sense that it recovers the numerical 
  response of the reference \ac{hf} model (values and avarage rates in time).}
  \label{fig:VLPresults}
\end{figure}

The numerical results in Fig.~\ref{fig:VLPresults} {and 
Tab.~\ref{tab:error_vda}} show that the \ac{vda}-\ac{pp} model is successfully 
able to recover the accuracy of the \ac{hf} model, at the same reduced 
computational cost of the \ac{htc}-\ac{pp} model. {In particular, average 
rates of temperature at initial build-up and quasi-steady state regime match 
pretty well;} the only mismatch is observed at the cooling phase, where the 
\ac{vda}-\ac{pp} model overestimates the thermal inertia of the system.

\begin{table}[!h]
    \centering
    \begin{tabular}{ | c | c | c | c | c | }
        \cline{2-5}
        \multicolumn{1}{c|}{} & \multicolumn{2}{c|}{\textbf{\ac{hf} model}} 
        & \multicolumn{2}{c|}{\textbf{experiments}} \\ \hline
        \textbf{Channel} & \textbf{MAE [C]} & \textbf{MRE [\%]} & \textbf{MAE 
        [C]} & \textbf{MRE [\%]} \\ \hline
        CH2 & 3.6 & 1.5 & 3.6 & 1.5 \\ \hline
        CH4 & 2.0 & 0.7 & 4.5 & 1.5 \\ \hline
        CH8 & 2.0 & 0.6 & 6.5 & 1.9 \\ \hline
    \end{tabular}
    \vspace{0.2cm}
    \caption{{\ac{mcam} experiment. Mean Absolute Error (MAE) and Mean 
    Relative Error (MRE) of the temperature evolution during the printing phase 
    between the \ac{vda}-\ac{pp} model and the reference \ac{hf} or the 
    measured data.}}
    \label{tab:error_vda}
\end{table}

\section{Conclusions}
\label{sec:conclusions}

This work introduces a new rationale for physics-based model reduction in 
the thermal \ac{fe} analysis of metal \ac{am} by powder-bed fusion methods, 
the Virtual Domain Approximation (\ac{vda}). In view of the locality of the 
process, it is reasonable to exclude regions of low physical relevance from the 
domain of analysis to reduce the size of the problem. However, lack of 
experimental data hinder proper estimation of heat loss through the neglected 
regions and rather simple approximations have been considered so far. By 
contrast, the \ac{vda} is thought to integrate the physics being neglected to 
evaluate heat loss through an excluded region. Inspired by existing methods in 
casting solidification, it consists in replacing the 3D \ac{fe} model at the 
low-relevance region (e.g. loose powder bed, building plate) with a 1D heat 
conduction problem. Following this, the 1D problem is discretized at 
convenience and reformulated as a temperature-dependent Robin-type boundary 
condition for the 3D model in the reduced domain. 

Using this approach, reduced models obtained by either excluding the powder 
bed or the powder bed and the building plate were derived and confronted with 
(1) a reference complete model and (2) the same reduced model, but with a 
constant boundary condition, i.e. constant \ac{htc} and environment 
temperature. As observed in the numerical experiments, the computational 
benefit of meshing a smaller geometry (the simulation time is reduced one 
order of magnitude) is accompanied by increased accuracy with respect to (2). 
In fact, the new method mostly recovers the thermal response predicted by (1) 
with the same computational cost of (2). Hence, this domain reduction strategy 
arises as an alternative that strikes good balance between efficiency and 
accuracy.

Even using reduced model variants, experimental validation and industrial 
applications are still challenged by high computational cost and uncertainty 
of the material and process parameters. To deal with this issue, this work 
turns to a methodology unprecedented in the field: an \ac{hpc} platform that 
brings together (1) FEMPAR-AM, a \ac{fe} model for the simulation of AM 
processes, designed to efficiently exploit distributed-memory supercomputers; 
and (2) Dakota, a suite of iterative mathematical and statistical methods for 
parametric exploration of computational models. As shown in the \ac{mcam} 
experiment, this advanced framework enables fast and automatized sensitivity 
analysis and parameter estimation. Hence, this kind of synergy may be useful 
in metal AM, not only to address verification and validation, but also for 
practical optimization problems.

One remaining question is to {study the method for builds with 
curved shapes. The \ac{vda} assumes unidimensional heat loss, normal to the 
discrete interface. This is optimal for flat surfaces, but for smooth ones, it 
may lead to an underestimation of heat loss that, in some cases, can be 
compensated with simple geometrical correction factors proposed in 
Sect.~\ref{sec:virtual}.}

Another interesting line of work is to push ahead the \ac{vda} capabilities, by 
reducing the domain up to the last simulated layers. In this scenario, the 
domain is thought as moving upwards to track the growth of the geometry, while 
keeping its size controlled either with a fixed or error-based criterion.


\section*{Acknowledgements}
\label{sec:acknowledgements}

Financial support from the EC - International Cooperation 
in Aeronautics with China (Horizon 2020) under the \emph{EMUSIC} project 
(\emph{Efficient Manufacturing for Aerospace Components USing Additive 
Manufacturing, Net Shape HIP and Investment Casting}), the EC - Factories of 
the Future (FoF) Programme under the \emph{CA}$\times $\emph{Man} Project 
(\emph{Computer Aided Technologies for Additive Manufacturing}) within 
\emph{Horizon 2020} Framework Programme and the Spanish 
Government-MINECO-Proyectos de I+D 
(Excelencia)-DPI2017-85998-P-ADaMANT-Computational Framework for Additive  
Manufacturing of Titanium Alloy are gratefully acknowledged. E. Neiva 
gratefully acknowledges the support received from the Catalan Government 
through a FI fellowship (2019 FI-B2-00090; 2018 FI-B1-00095; 2017 FI-B-00219). 
E. Salsi gratefully acknowledges the support received from the European Union’s 
Horizon 2020 research and innovation programme under the Marie Skłodowska-Curie 
Grant Agreement No. 746250. S. Badia gratefully acknowledges the support 
received from the Catalan Government through the ICREA Acad\`emia Research 
Program. Financial support to CIMNE via the CERCA Programme  / Generalitat de 
Catalunya is also acknowledged. The authors thankfully acknowledge the computer 
resources at TITANI by CaminsTECH, the support provided by O. Colomés, 
concerning the usage of Dakota, and the help from J. J. Moya in the numerical 
experiments. The experimental work is funded by the Science \& Industry 
Endowment Fund program RP04-153 \emph{Manufacturing a small demonstrator 
aero-engine entirely through additive manufacturing} and Australia Research 
Council IH130100008 \emph{Industrial Transformation Research Hub for 
Transforming Australia’s Manufacturing Industry through High Value Additive 
Manufacturing}, including financial support from Safran Power Units and Amaero 
Engineering.

\bibliographystyle{abbrvnat}
\bibliography{art028}

\end{document}